\documentclass[journal, doublecolumn, 10pt, a4paper]{IEEEtran}

\usepackage{amssymb}
\usepackage{amsmath}
\usepackage{amsthm}
\usepackage{mathrsfs}
\usepackage{verbatim}
\usepackage{bbm}
\usepackage{stmaryrd}
\usepackage{soulutf8}
\usepackage{cite}

\DeclareFontFamily{OT1}{pzc}{}
\DeclareFontShape{OT1}{pzc}{m}{it}{<-> s * [1.10] pzcmi7t}{}
\DeclareMathAlphabet{\mathpzc}{OT1}{pzc}{m}{it}

\ifCLASSINFOpdf

    \usepackage[pdftex]{graphicx}
    \DeclareGraphicsExtensions{.pdf,.jpeg,.png}
\else
	\usepackage[dvips]{graphicx}
	\DeclareGraphicsExtensions{.eps}
\fi
\ifCLASSOPTIONcompsoc
	\usepackage[caption=false,font=normalsize,labelfont=sf,textfont=sf]{subfig}
\else
	\usepackage[caption=false,font=footnotesize]{subfig}
\fi

\RequirePackage{lineno}

\usepackage{paralist}
\usepackage{enumerate}
\usepackage{enumitem}

\newlist{myitemize}{itemize}{3}
\setlist[myitemize]{noitemsep, topsep=0pt}
\setlist[myitemize,1]{label=\textbullet,leftmargin=1in}
\setlist[myitemize,2]{label=$\rightarrow$,leftmargin=1em}
\setlist[myitemize,3]{label=$\diamond$}

\usepackage{color}
\usepackage[usenames]{xcolor}

\usepackage{enumerate}
\usepackage{todonotes}
\usepackage{tabularx}

\usepackage{arydshln,leftidx,mathtools}
\usepackage{booktabs}

\usepackage{subfig}

\usepackage[euler]{textgreek}

\usepackage{tikz}
\usetikzlibrary{shapes,decorations}

\usepackage{multirow}
\usepackage{hhline}

\usepackage{array}
\newcolumntype{L}[1]{>{\raggedright\let\newline\\\arraybackslash\hspace{0pt}}m{#1}}
\newcolumntype{C}[1]{>{\centering\let\newline\\\arraybackslash\hspace{0pt}}m{#1}}
\newcolumntype{R}[1]{>{\raggedleft\let\newline\\\arraybackslash\hspace{0pt}}m{#1}}
\SetSymbolFont{stmry}{bold}{U}{stmry}{m}{n}

\makeatletter
\newcommand*{\Rom}[1]{\expandafter\@slowromancap\romannumeral #1@}

\makeatother

\DeclareMathOperator*{\argmax}{\text{argmax}}

\newcommand{\N}[0]{\ensuremath{\mathcal{N}}}
\newcommand{\GP}[0]{\ensuremath{\mathcal{GP}}}
\renewcommand{\vec}[1]{\ensuremath{\boldsymbol{#1}}}

\newtheorem*{theorem*}{Theorem}

\newtheorem*{definition*}{Definition}

\begin{document}

\title{RSS Models for Respiration Rate Monitoring}
\author{
	H\"{u}seyin~Yi\u{g}itler,  Ossi~Kaltiokallio, Roland~Hostettler, Riku~J\"{a}ntti, Neal~Patwari, and 
	Simo~S\"{a}rkk\"{a}
    \thanks{ H\"{u}seyin~Yi\u{g}itler,  Ossi~Kaltiokallio, and Riku~J\"{a}ntti are with Aalto University, 
    Department of Communications and Networking. e-mail: \{name.surname\}@aalto.fi}
	\thanks{Roland~Hostettler and  Simo~S\"{a}rkk\"{a} are with Aalto University, 
	Department of  Electrical Engineering and Automation. e-mail: \{name.surname\}@aalto.fi}
	\thanks{Neal~Patwari is with the Dept. of Electrical \& Computer Engineering, University of Utah, and Xandem 
	Technology LLC, Salt Lake City, UT 84112 USA (e-mail: npatwari@ece.utah.edu).}
	}
\maketitle

\begin{abstract}
Received signal strength based respiration rate monitoring is emerging as an alternative non-contact technology. These 
systems make use of the radio measurements of short-range commodity wireless devices, which vary due 
to the inhalation and exhalation motion of a person. The success of respiration rate estimation using such measurements 
depends on the signal-to-noise ratio, which alters with properties of the person and with the measurement system. To 
date, no model has been presented that allows evaluation of different deployments or system configurations for 
successful breathing rate estimation. In this paper, a received signal strength model for respiration rate monitoring 
is introduced. It is shown that measurements in linear and logarithmic scale have the same functional form, and the 
same estimation techniques can be used in both cases. The implications of the model are validated under varying 
signal-to-noise ratio conditions using the performances of three estimators: batch frequency estimator, recursive 
Bayesian estimator, and model based estimator. The results are in coherence with the findings, and they imply that 
different estimators are advantageous in different signal-to-noise ratio regimes.
\end{abstract}

\section{Introduction}

Serious respiratory disease can be identified by continuously monitoring variation of the respiration 
rate~\cite{Cretikos2008}. The importance of this vital sign is well acknowledged, and both contact and non-contact 
measurement systems are commercially available~\cite{AL-Khalidi2011}. Non-contact respiration rate monitoring is 
advantageous compared to contact systems in terms of improved patient's comfort and less patient distress, which result 
in improved accuracy. Although these devices fail to provide a measure to indicate actual respiratory 
gas-exchange~\cite{Folke2003}, patient's status can be attributed as stable or as at high-risk 
by continuously monitoring their respiration rate. In this paper, we consider non-contact respiration rate monitoring 
using commercially available low-cost standard wireless nodes. The main aim of these kind of systems is to estimate the 
respiration rate using the low-amplitude signal variations due to inhale and exhale motion.  

The impact of respiration on the received signal has a complex relationship with the geometry and electrical 
properties of the objects in the environment, which is here referred to as the \emph{radio channel}. As the granularity 
of the channel measurements increases, the information about the respiration can be better seen in the measurements. 
For example, high granularity measurements of different radar systems can be used for developing high quality 
monitoring systems~\cite{Greneker1996,Staderini2002,Adib2015}. However, higher quality channel 
measurements require complicated and expensive measurement systems. In contrast, monitoring systems based 
on commercially available narrowband communication systems are readily available, cheap, and easy to deploy. For 
example, several studies have evaluated the performance and different aspects of 
orthogonal-frequency-division-multiplexing (OFDM) based WiFi (IEEE 802.11 a, g, n, ac) \cite{Liu2014}, and low-power 
IEEE 802.15.4 compliant \cite{Kaltiokallio2014, Patwari2014a, Patwari2014} systems. In this paper, narrowband and 
low-power systems, which provide received signal strength (RSS) measurements to assess the state of the propagation 
channel, are considered. Such measurements are the most challenging for breathing monitoring since their measurements 
provide only coarse information about the channel. However, it is possible to build very low-cost systems using 
off-the-shelf components or using already available radios of mobile devices or smart appliances. 

Breathing monitoring using narrowband systems can be efficiently realized using a single pair of transmitter (TX) 
and receiver (RX) nodes~\cite{Kaltiokallio2014}. The RSS measurements of the RX can be modeled as a single tone 
sinusoid contaminated with noise~\cite{Patwari2014a}, whose maximum likelihood estimator using discrete time 
observations is known to be equivalent to the peak of discrete power spectral density of the 
measurements~\cite{Rife1974}. The estimation quality of such estimators, however, exhibits a thresholding behavior 
depending on the signal-to-noise ratio (SNR) of the measurements. When the SNR is low, the mean-square-error increases 
very rapidly and the estimation performance quickly degrades. To date, the SNR of RSS measurements for 
breathing rate monitoring has not been validated and no explanation of performance degradation has been provided. In 
this paper, a RSS model for respiration rate monitoring is presented, and its extensions for measurements in 
logarithmic scale and under quasi-linear movements are introduced. These allow 
one to predict the expected performance for a deployment scenario and patient position, and better assess the required 
system configuration for successful breathing monitoring.

In this paper, a RSS model of narrowband communication systems for respiration rate monitoring using the reflection 
based models is presented. The reflection model, previously explored for RSS-based localization \cite{Yigitler2017b, 
Kaltiokallio2017b}, is applied to respiration rate monitoring. Based on this model, first, it is shown that the 
breathing signal is frequency modulated into the RSS in linear scale due to small periodic movements. Such a signal 
exhibits relatively strong components on more than one frequency tones so that breathing estimators making use of this 
feature (see \cite{Hostettler2017}) perform better in high SNR conditions. Then, it is also shown that the RSS in 
logarithmic scale also has the same form, and an explicit model is derived. The model itself allows us to evaluate 
feasibility of different breathing rate estimation techniques by enabling SNR evaluation of any given deployment. The 
impact of several parameters are evaluated both numerically and empirically. This paper makes the following 
contributions:
\begin{myitemize}[leftmargin=3.5mm]
	\item A series expansion of the reflection-based RSS model is derived. This allows one to find several 
	approximations to the observed RSS variations.
	\item A RSS model for small periodic perturbations (e.g. breathing motion) is derived. It is shown 
	that the RSS variation due to such motions yields discrete tones at the harmonics of the perturbation frequency.
	\item A RSS model for small periodic perturbations when there is a linear movement is given. It is shown that the 
	movement itself modulates the periodic perturbation, making the estimation a more challenging problem.
	\item Based on the models, various scenarios observed in empirical data are discussed and their impact on the 
	observed RSS is shown.
	\item The performance of three different estimators are compared, and their performances are linked to the 
	implications of the model.   
\end{myitemize}

The remaining part of the paper is organized as follows. First, the related work is summarized in 
Section~\ref{sec:related-work}. The RSS measurement model and its series expansion are derived in 
Section~\ref{sec:models}. The impact of small periodic movements and 
linear movements are derived, and various numerical evaluations are presented in the same section. Breathing rate 
estimation techniques are introduced in Section~\ref{sec:estimation}, before giving empirical evaluations in 
Section~\ref{sec:experiments}. The conclusions are drawn in Section~\ref{sec:conclusion}.

\section{Related Work}\label{sec:related-work}
The importance of respiration rate has resulted in development of several respiration rate monitoring systems using 
different physical parameters (e.g. temperature, chest effort etc.) ~\cite{AL-Khalidi2011}. In this section, we only 
provide a brief review of radio-frequency based respiration rate monitoring methods, focusing on RSS-based approaches. 
The reader is referred to, for example, the works by AL-Khalidi et al.~\cite{AL-Khalidi2011} and Folke et 
al.~\cite{Folke2003} for comprehensive technological overviews. 

Respiration rate monitoring using radio frequency devices is a non-contact solution that has attracted significant 
attention. There are three different radar technologies that have 
been used for the purpose as has been reviewed by Li et al.~\cite{Li2013}. The first work that appeared in the 
literature uses continuous wave (CW) Doppler radar system. Several studies have been published to analyze different 
aspect of similar systems, which can also estimate the heart rate along with respiration rate~\cite{Salmi2012}. The 
impulse radio ultra wideband (IR-UWB) systems are also used for respiration rate monitoring~\cite{Staderini2002}. They 
radiate and consume little power, may coexist well with other instruments, and perform better in environments with 
interference and severe multipath~\cite{Lazaro2010}. The characteristics of the received signal of IR-UWB systems were 
investigated by Venkatesh et al.~\cite{Venkatesh2005}. However, these systems cannot cope with the impact of 
other motion or presence of more than one person~\cite{Adib2015}. Linear frequency modulated continuous wave (LFM-CW) 
systems can distinguish different reflector positions using their linearly varying frequency. This property has been 
used by Adib et al.\ for first estimating position of multiple persons in an environment~\cite{Adib2014}, and then 
estimating the vital signs of each individual~\cite{Adib2015}. In this work, we show that indeed RSS has similar 
characteristics as the measurements of radar solutions, but require more carefully adjusted deployments. 

The radar based solutions require a sophisticated hardware development for the vital sign monitoring. However, 
recent works on environmental sensing motivated RSS-based respiration rate monitoring using commodity wireless 
communication devices. The first work studying the feasibility of such systems makes use of multiple links formed 
by a mesh network of IEEE 802.15.4 nodes to estimate the breathing rate of a single person in the environment 
~\cite{Patwari2014a}.
Later, this system was extended to estimate the location of a breathing person~\cite{Patwari2014}. 
Several practical problems associated with the system are addressed in~\cite{Kaltiokallio2014} by using only one pair of
TX-RX nodes, detecting the moments breathing estimation is not possible, and using various signal 
processing techniques to improve SNR of the measurements. Due to widespread availability of WiFi, the communication 
channel  measurements of these systems have also been used for respiration monitoring~\cite{Liu2014, Abdelnasser2015}. 
The work by 
Abdelnasser et al.\ is based on the RSS measurements of WiFi systems~\cite{Abdelnasser2015}. On the other hand, the 
channel state information (CSI) output of OFDM-based WiFi radios provide higher granularity measurements of the 
communication channel, and have been used for vital sign monitoring purposes~\cite{Liu2015, 
Wang2016, Liu2016a}. The CSI contains a complex channel gain estimate at each sub-carrier. In case the transmitted 
power is constant, the amplitude variation of these gains define the RSS variation at each sub-carrier in linear scale. 
Therefore, it can be argued that the RSS-based models and algorithms can be applied directly to CSI based systems that 
use amplitudes of the complex CSI vector components. 

In this work, we model the RSS of narrowband communication systems for small periodic perturbations with unknown 
frequency, direction, and amplitude. It is shown that, similarly to signals in radar systems, the RF signal frequency 
modulate the periodic vital signals.  The resultant model shows explicit relation between the respiration signal and 
initial position of the person, her orientation with respect to link-line, amplitude of the breathing, and the 
wavelength of the communication system. The SNR depends on these parameters as well as the noise power which 
dictates the performance of single tone parameter estimation techniques~\cite{Rife1974}. This result is coherent with 
empirical results of Luong et al.~\cite{Luong2016}, real-time spectrum analyzer measurements in~\cite{Kaltiokallio2014},
and findings of Wang et al.~\cite{Wang2016}. The model also shows various interesting situations arising in practical 
deployments including the measurements showing only odd or even harmonics. It is further shown that the logarithmic 
transformation taking place in typical RSS measurement systems~\cite{Yigitler2017} does not change the signal type. 
Therefore, the developed model can be used for evaluating various deployment conditions that enable successful 
breathing rate monitoring.

\begin{figure}[!t]
	\centering
	\setlength{\tabcolsep}{0.3cm}
	\includegraphics[width=5cm]{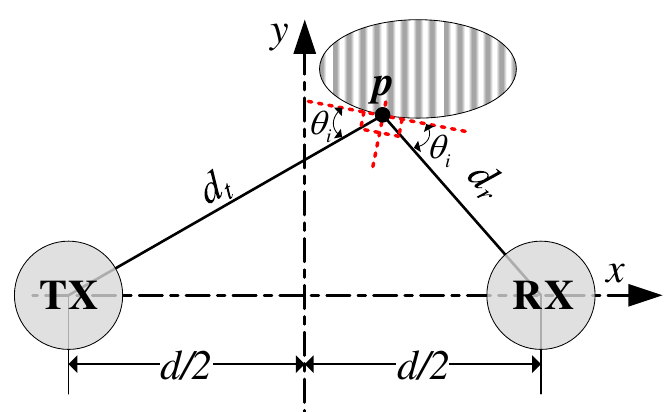} 
	\caption{A propagation scenario for RSS modeling and considered reference frame. The transmitter (TX) node placed 
	on point $\boldsymbol{p}_t$ emits a narrowband signal and the receiver (RX) node placed on $\boldsymbol{p}_r$ 
	receives the signal. Incident ray is reflected from point $\boldsymbol{p}$ on the surface of the object, which has 
	a distance 	$d_t= \|\boldsymbol{p}_t - \boldsymbol{p}\| \text{ m}$ to the TX and $d_r = \|\boldsymbol{p}_r - 
	\boldsymbol{p}\| \text{ m}$ to the RX. The incident ray has $\theta_i$ radians incidence angle. The TX and RX are 
	separated by $d = \|\boldsymbol{p}_t - \boldsymbol{p}_r\| \text{ m}$. } 
	\label{fig:geometry}
\end{figure}

\section{Received Signal Strength Models}\label{sec:models}
In this section, we first review the RSS model, and then derive temporal variations of the RSS in linear scale for an 
object\footnote{In this paper, we assume that the object is rigid so that any movement affects all points in its 
interior and on its boundary by the same amount.} making movements in the form 
\begin{equation}\label{eq:movement-dynamics}
\boldsymbol{p}(t) = \boldsymbol{p}_0 + \boldsymbol{v} t +  g(t) \boldsymbol{\delta},
\end{equation} 
where $\boldsymbol{p}_0$ is a reference point at $t=0$, $\boldsymbol{v}$ is the constant velocity with 
amplitude $v = \|\boldsymbol{v}\|$ for Euclidean norm $\|\cdot\|$, $t$ is the time elapsed since the epoch, 
$g(t)$ is the low-amplitude movement displacement from the initial position and $\|\boldsymbol{\delta}\| = 1$ is the 
constant movement direction. We start with small perturbation case, where  $v = 0$ and small amplitude $g(t)$ is 
periodic with frequency $f$. Then, we include quasi-linear constant velocity movements characterized by $v > 0$. The 
obtained models are extended to RSS measurements systems that perform logarithmic transformations and show that the 
output of these systems have similar form as in the linear scale case.

\subsection{Background}

For environmental sensing applications, the phenomenon of interest is estimated using the variation in RSS measurements 
compared to the measurement acquired when the object of interest is not in the medium, that is, the baseline RSS. A 
detailed analysis of the baseline RSS is provided in~\cite{Yigitler2017}, and it is concluded that if the complex 
channel gain $\alpha$ is constant, RSS is approximately a Gaussian random variable. This result also implies 
that if the medium is constant for the duration of acquiring the baseline RSS, their population mean converges 
to a constant $\mathcal{P}_r$ due to the \emph{strong law of large numbers} (SLLN). Then, the time average of the 
measurements also converges to the same constant since the channel is mean ergodic  when the environment is static. In 
other words, the baseline RSS $\mathcal{P}_r$  is given by
\begin{equation}
 \mathcal{P}_r = 10 \log_{10}(2\sigma^2+\varrho \sigma^2),
\end{equation}
where $\varrho$ is the signal-to-noise ratio (SNR) of the received signal under static channel conditions, and  
$\sigma^2$ is the noise variance of the zero mean Gaussian noise in signal samples. Therefore, a 
practical measurement model for environmental sensing applications is 
\begin{equation}\label{eq:rss-measurement}
r[k] \triangleq \mathcal{P}[k] - \mathcal{P}_r 
	\approx 10 \log_{10}\left( \frac{|\alpha[k]|^2}{|\alpha_r|^2}\right) + \nu[k],
\end{equation}
where $\alpha[k]$ is the channel gain when $k$\textsuperscript{th} RSS sample is acquired, and and $\alpha_r$ 
is the constant channel gain of the baseline RSS. The last term $\nu[k]$ is the joint noise process of all noise 
sources including white noise, round-off errors, quantization noise, and any source of uncertainty due to modeling 
errors, and its variance depends also on the current SNR value $\varrho[k]$, which is time varying for the non-static 
case.

The expression in Eq.~\eqref{eq:rss-measurement} implies that the complex channel gains $\alpha[k]$ and $\alpha_r$ 
define the measurements. For the scenario visualized in Fig.~\ref{fig:geometry}, the ratio of channel 
gains\footnote{The channel gain ratio follows from logarithmic scale definitions. If linear scale measurements are 
going to be 
used without calculating the amplitude ratio, the subsequent developments are valid through the relation 
$d^{-\eta/2}(R-1)$, which has the same spectral 
properties as $R$ but with different DC term and amplitude scale. See Eq.~\eqref{eq:R-linear-scale-cos-sine} and its
coefficients in Eq.~\eqref{eq:R-linear-scale-coefficients}.} has 
already been investigated in~\cite{Yigitler2017b, Kaltiokallio2017b}, and it has been shown that 
\begin{equation}\label{eq:reflection-model-ratio}
R \triangleq \frac{|\alpha|^2}{|\alpha_r|^2} = {1+G^2-2{G}\cos\left(\frac{2\pi 
\Delta}{\lambda}\right)},
\end{equation}
where $\Delta$ is excess path length traversed by the ray reflected from the object's surface, $G$ is 
the \emph{effective reflection coefficient}, $\lambda$ denotes the wavelength, and we have dropped sample index 
$[k]$ from $G$ and $\Delta$. The effective reflection coefficient is defined as
\begin{equation}\label{eq:G-definition}
G \triangleq \frac{\Gamma}{(1+\Delta/d)^{\eta/2}},
\end{equation}
where $\Gamma$ is the Fresnel reflection coefficient and $\eta$ is the path loss exponent modeling the fading 
experienced by both of the components \cite[ch.~4]{Rappaport2002}. The excess path length $\Delta$ parametrizes 
the ellipse tangent to the interacting object at the point $\boldsymbol{p}$ (cf. 
Fig.~\ref{fig:geometry}), and is defined as  
\begin{equation}\label{eq:Delta-definition}
\Delta \triangleq \|\boldsymbol{p} - \boldsymbol{p}_t\| + \|\boldsymbol{p} - \boldsymbol{p}_r\| - \|\boldsymbol{p}_r - 
\boldsymbol{p}_t\|= d_t + d_r - d,
\end{equation}
for the symbols visualized on Fig.~\ref{fig:geometry}. 

\subsection{Effect of Low Amplitude Periodic Perturbations}
In the previous section, we analyzed the variation of RSS when the object abruptly appears in position which yield 
$\Delta$ meters of excess path length when $\boldsymbol{p}_0$ is the reflection point shown in 
Fig.~\ref{fig:geometry}. Now, suppose the object makes a time varying movement in a constant direction 
$\boldsymbol{\delta}$, that is, the reflection point $\boldsymbol{p}_0$ moves to 
$
\boldsymbol{p}(t) = \boldsymbol{p}_0 + g(t) \boldsymbol{\delta}
$ 
at time instant $t$. At this new position, the excess path length can be found using its Taylor series expansion around 
$\boldsymbol{p} = 
\boldsymbol{p}_0$ as 
\begin{equation*}
\Delta(\boldsymbol{p}) = \Delta(\boldsymbol{p}_0) + g(t)(\nabla_{\boldsymbol{p}} \Delta)^\top \boldsymbol{\delta} + 
\mathcal{O}(g^2(t)),
\end{equation*}
where $\cdot^\top$ denotes the matrix transpose, $\nabla_{\boldsymbol{p}}$ is the gradient with respect to position 
$\boldsymbol{p}$, and we have used the fact that $\Delta$ is a smooth function 
of both coordinates except at $\boldsymbol{p} = \boldsymbol{p}_r$ or $\boldsymbol{p} = \boldsymbol{p}_t$. If the 
movement has a small amplitude $|g(t)| \ll 1$, the second and higher order terms in the Taylor series expansion can be 
ignored, and one can write 
\begin{equation}\label{eq:Delta-Taylor-series}
\Delta(\boldsymbol{p}) \approx \Delta_0 + g(t)\left[\frac{\boldsymbol{p}_0 - \boldsymbol{p}_{r} }{\| 
	\boldsymbol{p}_0 - \boldsymbol{p}_{r} \| } + 
\frac{\boldsymbol{p}_0 - \boldsymbol{p}_{t}}{\| \boldsymbol{p}_0 - \boldsymbol{p}_{t} \| 
}\right]^\top \boldsymbol{\delta},
\end{equation}
where $\Delta_0 =  \Delta(\boldsymbol{p}_0)$ and we have explicitly written the gradient of $\Delta$ with respect to 
$\boldsymbol{p}$ at $\boldsymbol{p} = 
\boldsymbol{p}_0$. Let us denote the inner product in Eq.~\eqref{eq:Delta-Taylor-series} as
\begin{equation}\label{eq:delta-Delta}
	\delta_{\Delta} \triangleq \left[
	\frac{\boldsymbol{p}_0 - \boldsymbol{p}_{r} }{\| \boldsymbol{p}_0 - \boldsymbol{p}_{r} \| } + 
	\frac{\boldsymbol{p}_0 - \boldsymbol{p}_{t}}{\| \boldsymbol{p}_0 - \boldsymbol{p}_{t} \| 
	}\right]^\top \boldsymbol{\delta}.
\end{equation}
Then, Eq.~\eqref{eq:reflection-model-ratio} can be written as 
\begin{equation}\label{eq:reflection-model-ratio-approx}
R \approx 1+{G}^2(\boldsymbol{p})-2G(\boldsymbol{p})
		\cos\left(2 \pi \frac{\delta_{\Delta}}{\lambda} g(t) + 2 \pi \frac{\Delta_0}{\lambda}  \right).
\end{equation}
It is to be noted that, for small amplitude perturbations 
satisfying $|g(t)\delta_{\Delta}| \ll d$, we have $G(\boldsymbol{p}) \approx G(\boldsymbol{p}_0)$. Consequently, the 
frequency of the cosine term in Eq.~\eqref{eq:reflection-model-ratio-approx} is defined by the perturbation amplitude 
$g(t)\delta_{\Delta}$, i.e., $R$ is a frequency modulated version of $g(t)$.

The amplitude ratio $R$ can be further simplified when $g(t)$ is a periodic function\footnote{The form of $g(t)$ is 
selected for simplicity. The analysis can be straightforwardly extended to any periodic function using their Fourier 
series expansion as it is shown in~\cite[ch.~5]{Carlson2002}.}. Suppose that $g(t) = A \sin(2 \pi f t)$, so that the 
\emph{affective amplitude of the periodic movement} $\tilde{A}$, and constant phase $\psi$ can be defined as 
\begin{equation}\label{eq:A-tilde-phi}
\tilde{A} \triangleq 2 \pi A \frac{\delta_\Delta}{\lambda}, \qquad  \psi \triangleq 2 \pi \frac{\Delta_0}{\lambda}. 
\end{equation}
Then, the Fourier series expansion of the cosine terms in Eq.~\eqref{eq:reflection-model-ratio-approx} are given by 
\begin{equation*}
\begin{aligned}
\cos\left(\tilde{A}  \sin(2\pi f t ) + \psi  \right) = \sum\limits_{m = 
	-\infty}^{\infty} J_m (\tilde{A} )\cos(2 \pi m f t + \psi),
\end{aligned}
\end{equation*} 
where $J_m(\cdot)$ is the Bessel function of the first kind~\cite[ch.~9]{Abramowitz1970}. Substituting this into 
Eq.~\eqref{eq:reflection-model-ratio-approx} yields 
\begin{equation}\label{eq:R-perturbation-series}
\begin{aligned}
{R}_1 \approx 1 +{G}^2 - 2 {G}
\sum\limits_{m = -\infty}^{\infty}J_m (\tilde{A})\cos(2 \pi m f t + \psi),
\end{aligned}
\end{equation}
which has the form of demodulated and low-pass filtered version of the respiration signal of IR-UWB systems derived by 
Venkatesh et al.~\cite{Venkatesh2005}. 

Using the properties of Bessel function, the expression in Eq.~\eqref{eq:R-perturbation-series} can be written in 
the form
\begin{equation}\label{eq:R-linear-scale-cos-sine}
\begin{aligned}
R_1 \approx c_0 +  \sum\limits_{m=1}^{\infty}\bigg(&c_{2m-1}\sin(2 \pi (2m-1) f t) \\ & +c_{2m}\cos(2 \pi 2 m f 
t)\bigg),
\end{aligned}
\end{equation}
where coefficients are given by
\begin{equation}\label{eq:R-linear-scale-coefficients}
c_m = 
\begin{cases}
1 +{G}^2 - 2 {G} J_0(\tilde{A})\cos(\psi), & m = 0, \\
~~~~~~~~~~~~4 G J_m(\tilde{A})\sin(\psi), & m \text{ is odd}, \\
~~~~~~~~~-4 G J_m(\tilde{A})\cos(\psi), & m \text{ is even}. 
\end{cases}
\end{equation}
\subsection{Effect of Quasi-linear Movements}

Let us suppose that the object shown in Fig.~\ref{fig:geometry} makes small periodic movements ($g(t)$) in addition to 
a constant velocity movement in another direction so that at time instant $t$, the initial point $\boldsymbol{p}_0$ 
moves to $
\boldsymbol{p}(t) = \boldsymbol{p}_0 + g(t) \boldsymbol{\delta} + \boldsymbol{v} t.
$
If $g(t)$ has a small amplitude and $t$ is close enough to the time epoch, the Taylor series expansion of the excess 
path length $\Delta$ is valid, and for this case Eq.~\eqref{eq:Delta-Taylor-series} can be written as
\begin{equation*}\label{eq:Delta-Taylor-series2}
\Delta(\boldsymbol{p}) - \Delta_0 \approx 
\left[\frac{\boldsymbol{p}_0 - \boldsymbol{p}_{r} }{\| 
	\boldsymbol{p}_0 - \boldsymbol{p}_{r} \| } + 
\frac{\boldsymbol{p}_0 - \boldsymbol{p}_{t}}{\| \boldsymbol{p}_0 - \boldsymbol{p}_{t} \| 
}\right]^\top \left[g(t) \boldsymbol{\delta} + \boldsymbol{v} t\right].
\end{equation*}
Similar to the definition in Eq.~\eqref{eq:delta-Delta}, let us denote the second inner product as
\begin{equation*}
\delta_v = \left[\frac{\boldsymbol{p}_0 - \boldsymbol{p}_{r} }{\| 
	\boldsymbol{p}_0 - \boldsymbol{p}_{r} \| } + 
\frac{\boldsymbol{p}_0 - \boldsymbol{p}_{t}}{\| \boldsymbol{p}_0 - \boldsymbol{p}_{t} \| 
}\right]^\top \boldsymbol{v},
\end{equation*} 
so that $\Delta(\boldsymbol{p}) - \Delta(\boldsymbol{p}_0) 
\approx g(t)\delta_\Delta + \delta_v t$. Then, Eq.~\eqref{eq:reflection-model-ratio} becomes  
\begin{equation}\label{eq:reflection-model-ratio-approx2}
\begin{aligned}
R(\boldsymbol{p}) \approx &1+{G}^2(\boldsymbol{p})- \\ & 2G(\boldsymbol{p})
\cos\left(2 \pi \frac{1}{\lambda} (\delta_{\Delta} g(t)+ \delta_v t) + \psi(\boldsymbol{p})  \right),
\end{aligned}
\end{equation}
where the position dependence of $R$ and phase $\psi$ defined in Eq.~\eqref{eq:A-tilde-phi}, which cannot be ignored 
for 
this case, are explicitly written. 

In case $g(t)$ is a sinusoidal in the form $g(t) = A \sin(2 \pi f t)$, as in the previous subsection, then 
Eq.~\eqref{eq:reflection-model-ratio-approx2} can be written as
\begin{equation}\label{eq:R-perturbation-linear-series}
\begin{aligned}
{R}_2 & \approx 1 +{G}^2(t) - \\ &2 {G}(t)
\sum\limits_{m = -\infty}^{\infty}J_m (\tilde{A})\cos\left(2 \pi \left(\frac{\delta_v}{\lambda} + m f 
\right)t+ \psi\right).
\end{aligned}
\end{equation}
Therefore, linear movements shift the frequency of the periodic movement and make the effective reflection coefficient 
$G$ defined in Eq.~\eqref{eq:G-definition} a time varying quantity since $\delta_v t / d$ term in 
$\Delta(\boldsymbol{p})$ cannot be neglected. 

It is to be noted that the expression in Eq.~\eqref{eq:R-perturbation-linear-series} has the same form as in 
Eq.~\eqref{eq:R-perturbation-series}, but the tones have ${\delta_v}/{\lambda} \text{ Hz}$ shifted frequency. 
The coefficients are the same as in Eq.~\eqref{eq:R-linear-scale-coefficients}.

\subsection{RSS in Logarithmic Scale}
Let us define the amplitude ratio $R$ in Eq.~\eqref{eq:reflection-model-ratio} in logarithmic scale as 
\begin{equation}\label{eq:reflection-model-ratio-logarithmic}
\begin{aligned}
\mathcal{R} \triangleq & 10 \log_{10}(e) \ln(R) \\ = &~~10 \log_{10}(e) \ln\left(1+G^2\right)  \\
	& + 10 \log_{10}(e) \ln\left( 1 - \kappa \cos\left(2 \pi \beta \Delta \right) \right),
\end{aligned}
\end{equation}
where $e$ is the base of the natural logarithm and we have defined
\begin{equation}\label{eq:series-defines}
	\beta \triangleq \frac{1}{\lambda}, \qquad \kappa \triangleq 2\frac{G}{1+G^2}.
\end{equation}

It can easily be verified that $G < 1$ for $\Gamma < 1$ so that $0 <\kappa < 1$ and 
the power series expansion of the second term in Eq.~\eqref{eq:series-defines} is given by
\begin{equation*}
\begin{aligned}
\ln \Big(1 -\kappa \cos(2 \pi \beta \Delta) \Big) &= -\sum\limits_{l=1}^{\infty}{\frac{1}{l}\big(\kappa \cos(2 \pi 
\beta \Delta )\big)^l} \\
&= b_0 + \sum\limits_{i=1}^{\infty}{b_i \cos(2 \pi i\Delta  \beta )},
\end{aligned}
\end{equation*}  
where the coefficients are given by
\begin{equation*}
\begin{aligned}
b_i = -\Delta \int\limits_{-\frac{1}{2\Delta}}^{\frac{1}{2\Delta}} {\sum\limits_{l=1}^{\infty}
	{\frac{\kappa^l}{l}\cos^l(2 \pi \Delta \beta)} \cos(2 \pi i \Delta \beta) d\beta},
\end{aligned}
\end{equation*}
for all $i \in \{0, 1, 2, \dots \}$. In addition, the cosine powers can be expanded as harmonics, 
\begin{equation*}
\begin{aligned}
\cos^l(\phi) = 
\begin{cases}
\frac{2}{2^l}\sum\limits_{i=0}^{\frac{l-1}{2}}{\binom{l}{i}\cos\big( (l-2 i) \phi\big)}, & l \text{ odd}, \\
\frac{1}{2^l}\binom{l}{\frac{l}{2}} + \frac{2}{2^l}\sum\limits_{i=0}^{\frac{l-2}{2}}{\binom{l}{i}\cos\big((l-2 i) 
\phi\big)}, & l \text{ even},
\end{cases}
\end{aligned}
\end{equation*}
where $\phi$ is an arbitrary argument of $\cos(\cdot)$. Due to orthogonality of the sinusoidal functions, $b_i$ are 
polynomials of $\kappa$ which can be written as
\begin{equation*}
\begin{aligned}
b_i = 
\begin{cases}
- ~\sum\limits_{l=1}^{\infty}{\frac{\kappa^{2 l}}{2 l 2^{2 l}}}\binom{2 l}{l}, & i = 0, \\
- \sum\limits_{l = \frac{i+1}{2}}^{\infty}{\frac{2 \kappa^{2 l - 1}}{(2 l - 1) 2^{2 l - 1}}}\binom{2 l - 
1}{\frac{2 l - i -1}{2}}, & i \text{ odd}, \\
- ~\sum\limits_{l = \frac{i}{2}}^{\infty}{\frac{2 \kappa^{2 l }}{(2 l ) 2^{2 l}}}\binom{2 l}{\frac{2 l - 
i}{2}}, & i \text{ even}. 
\end{cases}
\end{aligned}
\end{equation*}
Note that $b_i < 0$ for all $i=0,1,2, \cdots$, and for $\kappa<1$ and $i > 0$ we have $|b_i|>|b_{i+1}|$. The partial 
sums of  the coefficients are convergent, and after simplification and substituting definition of $\kappa$ in 
Eq.~\eqref{eq:series-defines} into this result yields 
\begin{equation*}
b_i = 
\begin{cases}
- \ln \left(1+G^2\right), & i = 0, \\
- 2\frac{G^i}{i},  & i >0 .
\end{cases}
\end{equation*} 
One important consequence is that $b_0$ is equal to the additive inverse of the first term in
Eq.~\eqref{eq:reflection-model-ratio-logarithmic}, and they cancel out. Therefore, the RSS measurement model in 
Eq.~\eqref{eq:reflection-model-ratio-logarithmic} can be written as
\begin{equation}\label{eq:reflection-model-ratio-logarithmic-series}
	\mathcal{R} = -20 \log_{10}(e)\sum\limits_{i=1}^{\infty}{ \frac{G^i}{i}  \cos\left(2 \pi \frac{i}{\lambda} 
		\Delta\right)}.
\end{equation}

The series in Eq.~\eqref{eq:reflection-model-ratio-logarithmic-series} implies that the periodic sinusoidal 
perturbation in Eq.~\eqref{eq:R-perturbation-series} after logarithmic transformation reads as
\begin{equation}\label{eq:perturbed-rss2}
\begin{aligned}
\mathcal{R}_1 \approx -20 \log_{10}(e) &\sum\limits_{m=-\infty}^{\infty} \sum\limits_{i=1}^{\infty}\Bigg\{ J_m (i 
\tilde{A}) \frac{G^i}{i} \\ &\cos(2 \pi m f t + i\psi)\Bigg\},
\end{aligned}
\end{equation}
where $\psi = 2 \pi \Delta(\boldsymbol{p}_0)/ \lambda$. Similarly, the impact of quasi-linear movement given in 
Eq.~\eqref{eq:R-perturbation-linear-series} for logarithmic scale is given by
\begin{equation}\label{eq:perturbed-rss3}
\begin{aligned}
\mathcal{R}_2 \approx -20 \log_{10}(e) &\sum\limits_{m=-\infty}^{\infty} \sum\limits_{i=1}^{\infty} \Bigg\{J_m(i 
\tilde{A}) \frac{G^i(t)}{i} \\ &  \cos\left(2 \pi \left(i\frac{\delta_v}{\lambda}+ m f \right) t + i \psi\right)\Bigg\}.
\end{aligned}
\end{equation}

The series in Eq.~\eqref{eq:perturbed-rss2} can be written in the form in 
Eq.~\eqref{eq:R-linear-scale-cos-sine}, 
\begin{equation}\label{eq:R-log-scale-cos-sine}
\begin{aligned}
\mathcal{R}_1 \approx \mathpzc{c}_0 + \sum\limits_{m=1}^{\infty}\bigg(&\mathpzc{c}_{2m-1}\sin(2 \pi (2m-1) f t) \\
&+ \mathpzc{c}_{2m}\cos(2 \pi 2 m f t)\bigg),
\end{aligned}
\end{equation}
where, for this case, the coefficients are given by
\begin{equation}\label{eq:R-log-scale-coefficients}
\mathpzc{c}_m = 
\begin{cases}
-20 \log_{10}(e)\sum\limits_{i=1}^{\infty}J_0(i\tilde{A})\frac{G^i}{i}\cos(i\psi), & m = 0, \\
~~40 \log_{10}(e)\sum\limits_{i=1}^{\infty}J_m(i\tilde{A})\frac{G^i}{i}\sin(i\psi), & m \text{ odd}, \\
-40 \log_{10}(e)\sum\limits_{i=1}^{\infty}J_m(i\tilde{A})\frac{G^i}{i}\cos(i\psi), & m \text{ even}. 
\end{cases}
\end{equation}
Similar to Eq.~\eqref{eq:R-perturbation-linear-series}, Eq.~\eqref{eq:perturbed-rss3} can be written in the same form 
with coefficients in Eq.~\eqref{eq:R-log-scale-coefficients}.

\begin{figure*}
\centering
\setlength{\tabcolsep}{0pt}
\begin{tabular}{ccc}
	\subfloat[]{\includegraphics[width=0.34\textwidth]{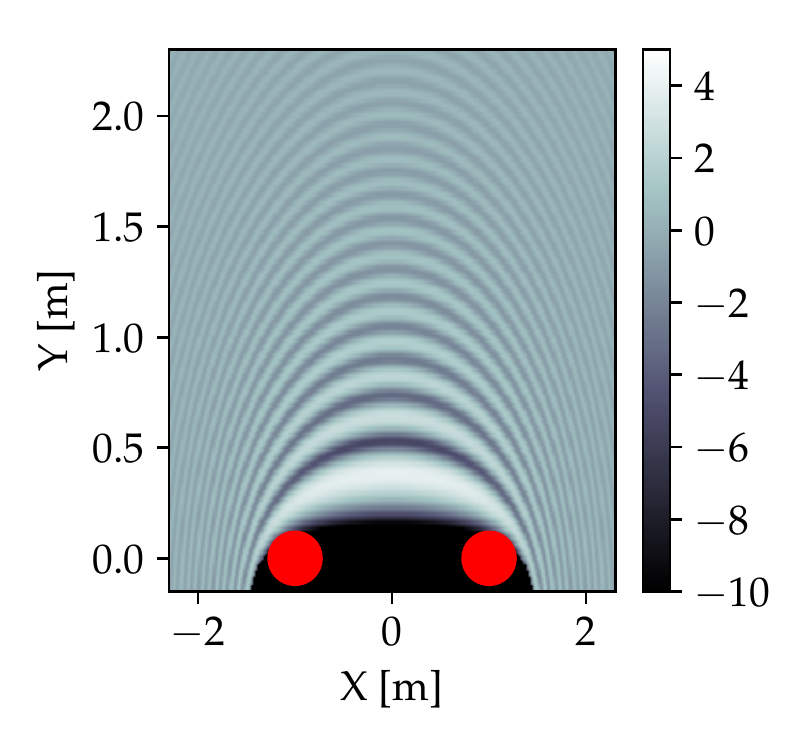} \label{fig:RSS-link}} &
	\subfloat[]{\includegraphics[width=0.32\textwidth]{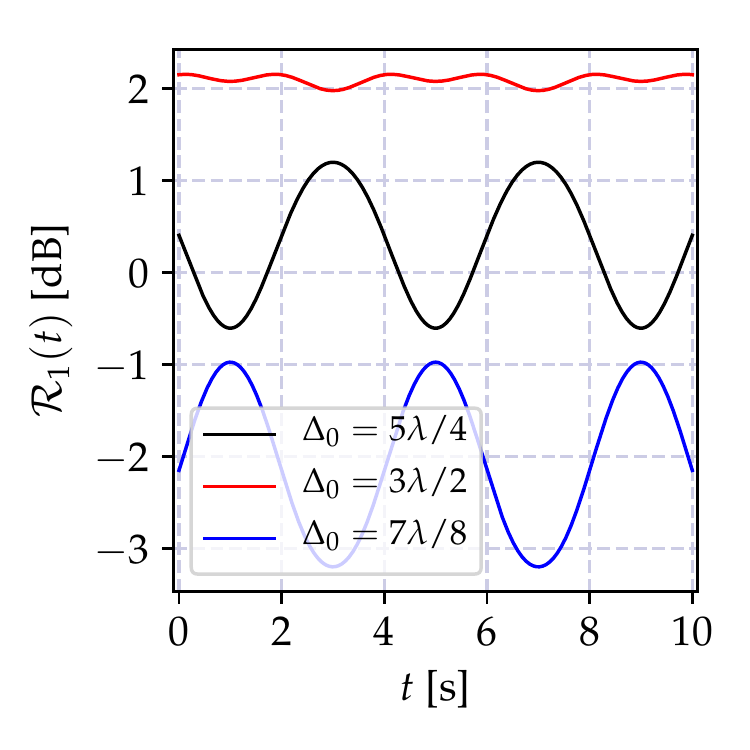} \label{fig:RSS-t}} &
	\subfloat[]{\includegraphics[width=0.32\textwidth]{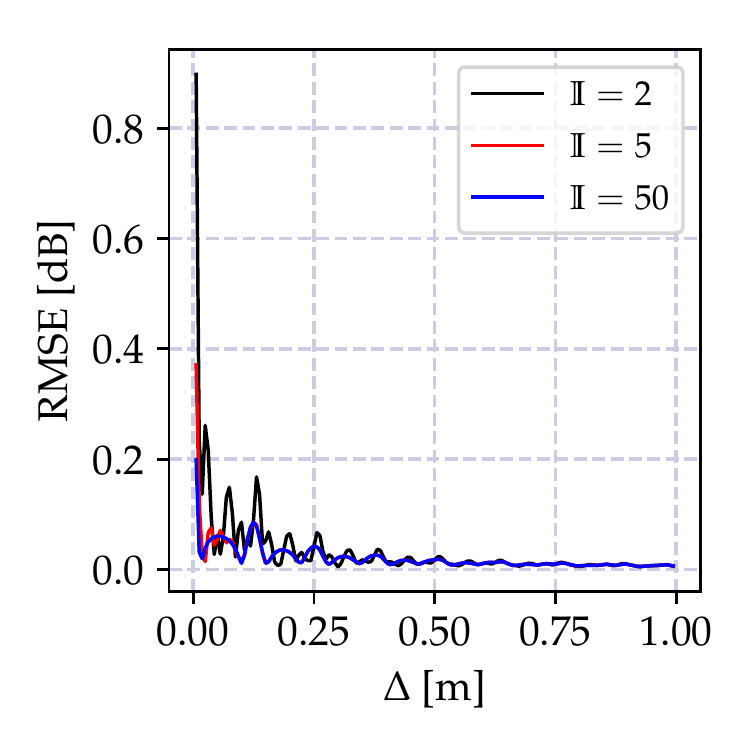} \label{fig:RMSE}} 
\end{tabular}
\caption{The RSS variation of a link with TX node at $\boldsymbol{p}_t = [-1, 0]^\top$ and RX node at 
$\boldsymbol{p}_r = [1, 0]^\top$ operating with wavelength $\lambda = 0.125 \text{ m}$ in an environment with path-loss 
exponent $\eta = 2$ when a circular object of radius $0.3 \text{ m}$ and relative permittivity $\epsilon_r = 1.5$ moves 
to different positions. In (a), the variation of RSS with object's position. In (b), the variation of RSS when the 
object perturbs the RSS with a sinusoidal movement in direction $\boldsymbol{\delta} = [0, -1]^\top$ and amplitude 
$A = 0.01$, and the object's initial positions yield specified excess path lengths $\Delta_0$. In (c), the variation of 
RMS error between the model output in Eq.~\eqref{eq:reflection-model-ratio-logarithmic} and approximation using two 
harmonics and $\mathbb{I}$\textsuperscript{th} order series in coefficients in Eq.~\eqref{eq:R-log-scale-coefficients}. 
}
\label{fig:RSS}
\end{figure*}

\subsection{Discussion}\label{sec:discussion}
Suppose that the periodic movement represents respiration of a person, which is monitored by a TX-RX pair operating at 
$2.4 \text{ GHz}$ ISM band so that $\lambda = 0.125 \text{ m}$. Although the respiration has a non-trivial relation 
between sex, age, and posture~\cite{Kaneko2012}, on average, it is a small quantity, for example, $A = 1 \text{ 
centimeter}$
maximum displacement\footnote{This is an example value, and may not correspond to any specific combination of sex, 
age, and posture for a chest or abdominal movement associated with the respiration~\cite{Kaneko2012}. }. 
An example deployment of such a system is expected to yield RSS measurements shown in Fig.~\ref{fig:RSS}.

\begin{figure}
\centering
\setlength{\tabcolsep}{0pt}
\begin{tabular}{cc}
	\subfloat[]{\includegraphics[width=0.25\textwidth]{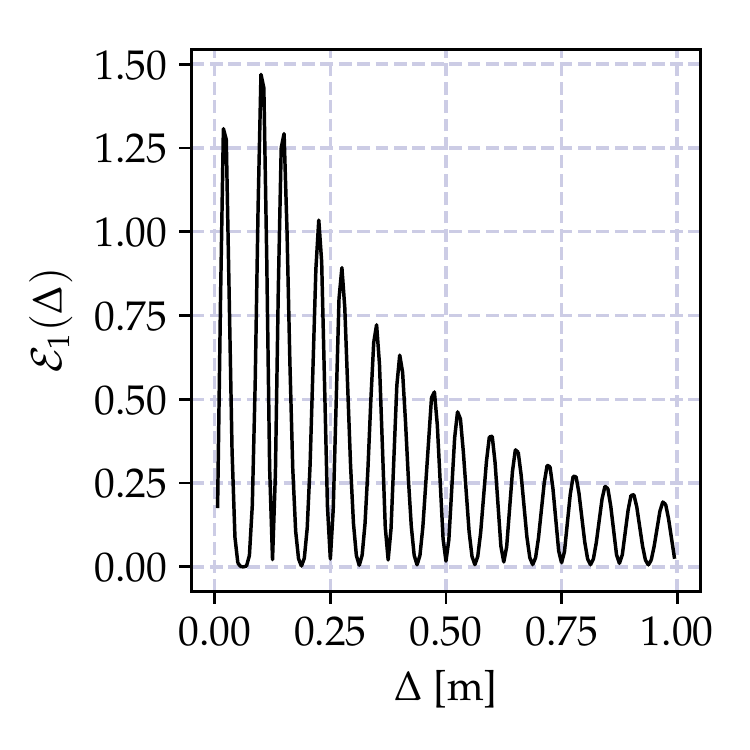} \label{fig:Energy}} &
	\subfloat[]{\includegraphics[width=0.25\textwidth]{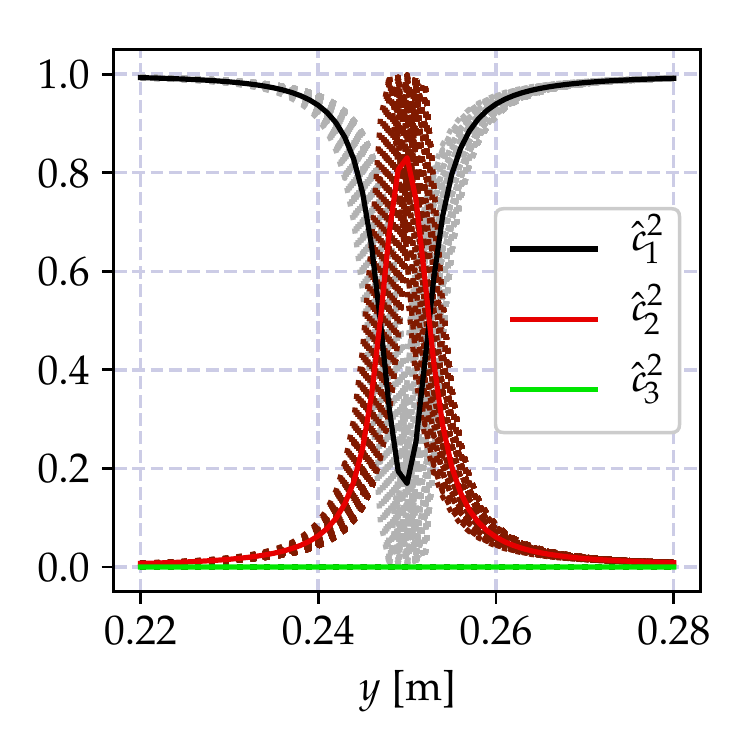} \label{fig:harmonic_energy}} 
\end{tabular}
\caption{In (a), the variation of signal energy $\mathcal{E}_1$ with excess path length $\Delta$. In (b), the relative 
contribution of harmonics to the signal energy for the object in Fig.~\ref{fig:RSS} moving on the mid-line between TX 
and RX ($x=0$) and on the specified $y$-axis values. In the plots, the coefficients are normalized with 
$\hat{\mathpzc{c}}_m^2 = {\mathpzc{c}_m^2}\big/{\mathcal{E}_1}$. Dots are coefficient square values for $16$ frequency 
channels separated by $5 \text{ MHz}$, and solid lines are their means.}
\label{fig:some_fig}
\end{figure}

A closer look at the coefficients in Eq.~\eqref{eq:R-log-scale-coefficients} reveals that when $\Delta \approx n 
\lambda/2$ for any integer $n$, all odd order harmonics $\mathpzc{c}_{2m-1}$ get closer to $0$ due to $\sin(n \pi)$ 
product. Similarly, when $\Delta \approx n \lambda/4$ for odd $n$, all even order harmonics $\mathpzc{c}_{2m}$ 
get closer to $0$. Therefore, for some 
special values of initial excess path value $\Delta_0$, the breathing of a stationary person may 
exhibit only even or odd harmonics in their measurement signal. The same argument is true for RSS 
measurements in linear scale (see Eq.~\eqref{eq:R-linear-scale-coefficients}), and it is independent of the actual 
periodic breathing signal function. In Fig.~\ref{fig:RSS-t}, Eq.~\eqref{eq:reflection-model-ratio-logarithmic} is used 
for calculating the RSS values for different $\Delta_0$ values. It can be seen that when $\Delta_0 = 3\lambda/2$, the 
RSS has a smaller amplitude but double perturbation frequency. At these distances, it can be observed from 
Fig.~\ref{fig:Energy} that the signal energy is very small. 

The respiration signal in Eq.~\eqref{eq:R-linear-scale-cos-sine} (and Eq.~\eqref{eq:R-log-scale-cos-sine}), is series 
expansion of a frequency modulated signal. For frequency modulated signals, the number of terms in their series 
expansion can found by Carson's bandwidth rule~\cite[ch.~5]{Carlson2002}, which states that it is enough to consider 
only $2(\tilde{A}/2 \pi + f)$ harmonics of the Fourier series in Eq.~\eqref{eq:R-perturbation-series} and 
Eq.~\eqref{eq:R-perturbation-linear-series}. Since the maximum displacement due to respiration can be assumed to 
satisfy $A = 0.01 \text{ m}$, and its frequency can be assumed to be less than $30$ breaths per minute ($0.5 \text{ 
Hz}$), as it has been done earlier~\cite{Patwari2014a, Kaltiokallio2014}, Carson's rule imply that only two harmonics 
are needed to represent most of the signal. Consequently, a conservative approximation of the 
amplitude ratio in linear scale $R$ and in logarithmic scale $\mathcal{R}$ is obtained by truncating the series in 
Eq.~\eqref{eq:R-linear-scale-cos-sine} at the second term.  For lower respiration rates, it is possible to truncate the 
series after the first term and obtain a single tone approximation which was used in previous 
works~\cite{Patwari2014a}.

The infinite series in Eq.~\eqref{eq:reflection-model-ratio-logarithmic-series} can be truncated at lower orders if the 
total signal energy is concentrated greatly in lower order harmonics. For this purpose, one can invoke Parseval's 
Theorem to find the signal energy of $\mathcal{R}$ from the periodic series expansion in 
Eq.~\eqref{eq:reflection-model-ratio-logarithmic-series} as 
\begin{equation*}
	\mathcal{E}_{\mathcal{R}} \triangleq \sum\limits_{i=1}^{\infty}\frac{G^{2i}}{i^2} = \mathrm{Li}_2(G^2),
\end{equation*}
where $\mathrm{Li}_2(\cdot)$ is the \emph{di-logarithm} function~\cite{Maximon2003}. Then, the total signal power is 
monotonically increasing with $G$. Since $G$, defined in Eq.~\eqref{eq:G-definition}, is decreasing function of 
$\Delta$ and increasing function of $\Gamma$, it attains its maximum when $\Delta = 0$ so that $\mathrm{Li}_2(G^2) \le 
\mathrm{Li}_2(\Gamma^2)$. This implies that the signal energy concentrated at lower order harmonics if $\Gamma$ is 
lower than $1$, which is defined by the incidence angle $\theta_i$ (cf. Fig.~\ref{fig:geometry}) and the object's 
relative permittivity. When the reflector object is close to the link line, $\theta_i$ approaches to $0$ radians making 
$\Gamma$ very close to $1$. In this case, high number of terms are needed to reach a good approximation of the series 
in Eq.~\eqref{eq:R-log-scale-coefficients}. However, when the object is sufficiently far away from the 
link-line, $\Gamma$ assumes smaller values and $\Delta$ increases, making $G$ a small quantity. When $G$ is smaller 
than $0.7$, first two harmonics in Eq.~\eqref{eq:reflection-model-ratio-logarithmic-series} contain $96.76 \%$ of total 
signal energy $\mathcal{E}_{\mathcal{R}}$, making two term approximation a reasonable choice. The 
variation of root-mean-square error (RMSE) with excess path length $\Delta$ is depicted in Fig.~\ref{fig:RMSE} for 
different truncation orders of the series in Eq.~\eqref{eq:R-log-scale-coefficients} while using only first two 
harmonics in 
Eq.~\eqref{eq:R-log-scale-cos-sine}. As shown, even for small $\Delta$, the two term series truncation yields small 
error.  

The observed breathing signal amplitude depends on the effective amplitude of the breathing displacement $\tilde{A}$ 
defined in Eq.~\eqref{eq:a-hat}, effective reflection coefficient $G$ defined in Eq.~\eqref{eq:G-definition}, and the 
excess path length when the person is in the initial position $\boldsymbol{p}_0$. 
The effect of position displacement on $\theta_i$ can be seen by inspecting its definition shown in 
Fig.~\ref{fig:geometry}, which yields 
\begin{equation}
\begin{aligned}
\theta_i &= \frac{\pi}{2} - \frac{1}{2}\arccos(p),\\
p(\boldsymbol{p}) &\triangleq \left(\frac{\boldsymbol{p} - \boldsymbol{p}_{r} }{\| \boldsymbol{p} - \boldsymbol{p}_{r} 
\| }\right)^\top 
\left(\frac{\boldsymbol{p} - \boldsymbol{p}_{t} }{\| \boldsymbol{p} - \boldsymbol{p}_{t} \| }\right).
\end{aligned}
\end{equation}
Then, it follows from the definition of Fresnel reflection coefficient that its first order Taylor series expansion 
reads as
\begin{equation}
\Gamma \approx \Gamma_0 - \Gamma_0\frac{p(\boldsymbol{p}) - p(\boldsymbol{p}_0)}{\sqrt{(p^2(\boldsymbol{p}_0) - 1) + 
2(p(\boldsymbol{p}_0) + 1)\epsilon_r}},
\end{equation}
where $\Gamma_0$ is the coefficient at initial position $\boldsymbol{p}_0$, and $\epsilon_r$ is the relative 
permittivity. When the object is between the nodes, and as it gets closer to the link-line $p$ gets closer to $-1$,  
the rate of change of $\Gamma$ increases so that small amplitude perturbations may significantly alter $\Gamma$. 

Consider the signal energy of the approximate (two harmonics and two terms in coefficient series) signal, which is 
given by
\begin{equation}\label{eq:Energy-log-scale}
 \mathcal{E}_1 \triangleq \mathpzc{c}_1^2 + \mathpzc{c}_2^2. 
\end{equation}
Its variation with $\Delta$ for the scenario used in generating the results in Fig.~\ref{fig:RSS} is depicted in 
Fig.~\ref{fig:Energy}. As shown, the energy may significantly change with the object's position only. One consequence 
of this result is there are certain positions where small changes drastically degrades the SNR. For example, 
when the object in Fig.~\ref{fig:RSS} is moving on the mid-line between TX and RX nodes, at certain $y$-axis values, 
the relative importance of the second harmonic exceeds the first harmonic as shown in Fig.~\ref{fig:harmonic_energy}. 
When excess path length is close to such a value, even small wavelength variations may change SNR of the measurements 
as can be observed from the dot plots in the figure. Furthermore, any uncertainty in electrical parameters of the 
object or its geometry increases the uncertainty further. Therefore, the SNR is statistical, and it is a very tedious 
task to accurately compare the model output with measurement data.

The analysis in the previous subsection and discussions above imply that:
\begin{enumerate}[leftmargin=3.5mm, label=\roman*.]
 \item The RSS measurements in linear scale Eq.~\eqref{eq:R-linear-scale-cos-sine} and logarithmic scale 
 Eq.~\eqref{eq:R-log-scale-cos-sine} have 
 similar forms.
 \item When the object moves even with a constant velocity, the perturbation spectrum shifts depending on the speed and 
direction of the movement. In this case, the perturbation 
frequency estimate has a different nature and it is required to first estimate the center frequency $\delta_v/\lambda$ 
(see Eq.~\eqref{eq:perturbed-rss3}). The other option is to stop estimating breathing rate when a movement is detected, 
as it has been done in~\cite{Kaltiokallio2014}.
 \item The signal energy can be estimated using Eq.~\eqref{eq:Energy-log-scale} as the higher order terms  
introduce small modeling error.  
 \item  The breathing signal's energy has a non-trivial relation 
 with breathing direction with respect to the link-line, breathing amplitude, initial position of the person, her 
 geometry, and electrical properties of her clothes. Therefore, the SNR of the breathing signal is 
 statistical, and some diversity mechanism is needed to improve breathing estimation quality. For example in 
~\cite{Kaltiokallio2014} frequency diversity and in~\cite{Patwari2014a} spatial diversity are used. In works using CSI 
output  of WiFi devices, the channel gains are estimated for large number of subcarriers, which yields improved 
breathing rate estimation quality. 
\end{enumerate}

\begin{figure}
\centering
\setlength{\tabcolsep}{0pt}
\begin{tabular}{c}
	\subfloat[]{\includegraphics[width=0.29\textwidth]{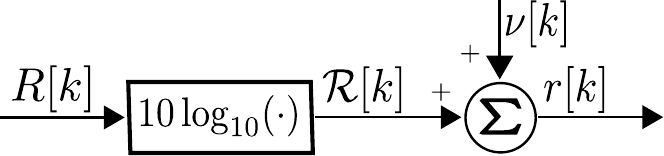} \label{fig:RSS-model-symbols}} \\
	\subfloat[]{\includegraphics[width=0.27\textwidth]{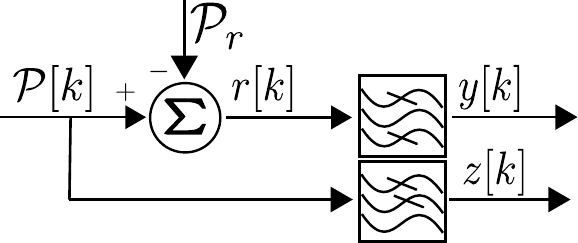} \label{fig:preprocessing}}
\end{tabular}
\caption{In (a), relations between various parameters introduced in Sec.~\ref{sec:models}, and in (b), preprocessing of 
the RSS measurements }
\end{figure}

\section{Breathing Rate Estimation}\label{sec:estimation} 
The analysis in the previous section conclude that the effect of respiration on RSS can be modeled in terms of the 
amplitude ratio in linear scale $R$ defined in Eq.~\eqref{eq:reflection-model-ratio} and in logarithmic scale 
$\mathcal{R}$ defined in Eq.~\eqref{eq:reflection-model-ratio-logarithmic} along with the additive noise $\nu$ as shown 
in Fig.~\ref{fig:RSS-model-symbols}. 

In this section, we discuss three approaches to estimate the respiration rate $f$. The estimators given in this section 
do not use the RSS measurements $r[k]$ directly, but its bandpass filtered version $y[k]$ or low-pass filtered 
$\mathcal{P}[k]$,
denoted as $z[k]$. These processing steps are visualized in Fig.~\ref{fig:preprocessing}. The breathing rate estimation 
problem can be casted as single tone parameter estimation of a deterministic sinusoid using discrete observations as 
has been done in the related works~\cite{Kaltiokallio2014, Patwari2014, Patwari2014a}. The problem has 
been extensively studied due to its importance in various application areas~\cite[ch.~13]{Kay1988}. In the first 
subsection, we give an overview of the batch-based frequency estimation technique. Then, we discuss a Bayesian 
formulation of the same approach, and summarize its recursive solution from the work by Qi et al.~\cite{Qi2002}. 
Finally, we review a recently introduced model-based approach~\cite{Hostettler2017}.

\subsection{Batch Spectral Analysis}
\label{sec:estimation:windowed}
The most straight forward and standard approach in the literature is to use spectral analysis using a batch of
measurement data. It is well established that the maximum likelihood estimate of the frequency of an unknown sinusoid 
is the frequency of the peak of the periodogram~\cite{Rife1974}. The peak of the periodogram can be calculated using 
the fast Fourier transform (FFT). However, this method requires significant frequency resolution and high SNR for
good performance since it is subject to thresholding effects. Here, we summarize the DFT-based method in a practical 
perspective. 

For the DFT-based method, the maximum and minimum breathing rates (perturbation frequencies) can be used for bandpass 
filtering the measurements $r[k]$ as shown in Fig.~\ref{fig:preprocessing}. Then, the output signal $y[k]$ is 
split into $M$ windows of length $N_w$ with overlap $N_o$ such that the data for the $m$\textsuperscript{th} window 
(for $m = 1, \dots, M$) is given by
\begin{equation}
    \vec{y}_{m}
        = \begin{bmatrix} y[n + 1] & y_j[n + 2] & \dots & y[n + N_w] \end{bmatrix}^\top,
    \label{eq:estimation:windowed:data}
\end{equation}
where the offset $n = (m-1) (N_w - N_o)$. Then, from Eq.~\eqref{eq:estimation:windowed:data}, the power spectral 
density (PSD) is estimated as
\begin{equation}
    \vec{S}_{m}[l] = |\vec{Y}_{m}[l]|^2,
    \label{eq:estimation:windowed:psd}
\end{equation}
where $\vec{Y}_{m}[l]$ is the DFT of $\vec{y}_{m}$ and the $l$\textsuperscript{th} frequency is given by
\begin{equation}
    f_l = \frac{l f_s}{N_w},
\end{equation}
for $l = 0, \dots, N_w-1$ and where $f_s$ is the sampling frequency.

Eq.~\eqref{eq:estimation:windowed:psd} naturally yields the whole spectrum for all frequencies $f_l$. In order to 
obtain a point estimate $\hat{f}$, the frequency corresponding to the maximum of the PSD is chosen, that is
\begin{equation}
        \hat{l} = \argmax_{l} \vec{S}_{m}[l], \qquad
        \hat{f}_\text{DFT} = \frac{\hat{l} f_s}{N_w},
    \label{eq:estimation:windowed:point_estimate}
\end{equation}
excluding the DC component.


\subsection{Recursive Bayesian Spectral Analysis}
\label{sec:estimation:recursive_bayesian}
Rather than processing entire overlapping windows at a time, the spectrum can also be estimated by using a recursive 
Bayesian spectrum estimation approach~\cite{Qi2002}. This method can cope with the DC term since it is possible to 
choose the resolution and sampling rate.  However, once these parameters are selected, the measurements 
must be low-pass filtered in order to keep the dimension of the state low. In other words, low-pass filtered RSS 
data $z[k]$ (see Fig.~\ref{fig:preprocessing}) can be used for estimating the perturbation frequency.

The Bayesian recursive method starts with writing $z[k]$ in terms of its Fourier series expansion as
\begin{equation}
    \begin{aligned}
        z[k]
            = a_0[k] + \sum\limits_{n = 1}^{N_{\text{KF}}} & a_n[k] \sin(2\pi f_n t_k) \\
            & + b_n[k] \cos(2\pi f_n t_k) + \tilde{\nu}[k]
    \end{aligned}
    \label{eq:estimation:recursive_bayesian:fourier_series_expansion}
\end{equation}
where $N_\text{KF}$ is the number of frequency bins, $a_0[k]$, $a_n[k]$, and $b_n[k]$ are the time-varying Fourier 
coefficients with slight abuse of notation, $f_n$ are the frequency bins, and 
$\tilde{\nu}[k] \sim \N(0, \sigma_\nu^2)$ is the measurement noise which is a filtered version of ${\nu}[k]$, but still 
assumed to be white.

Let $\vec{x}[k] = \begin{bmatrix} a_0[k] & \dots & a_N[k] & b_1[k] & \dots & b_N[k] \end{bmatrix}^\top$ be the vector 
of the Fourier coefficients, which are assumed to evolve as a Gaussian random walk according to
\begin{equation}
    \vec{x}[k] = \vec{x}[k-1] + \vec{w}[k]
    \label{eq:estimation:recursive_bayesian:coefficient_dynamics}
\end{equation}
where $\vec{w}[k] \sim \N(0, \boldsymbol{C}_w)$ is the process noise with covariance matrix $\boldsymbol{C}_w$. 
Furthermore, assume that the initial state $\vec{x}[k]$ is 
distributed according to $\vec{x}[0] \sim \N(\boldsymbol{m}_0, \boldsymbol{P}_0)$.

Combining Eq.~\eqref{eq:estimation:recursive_bayesian:fourier_series_expansion} and 
Eq.~\eqref{eq:estimation:recursive_bayesian:coefficient_dynamics},
 the following linear state space model is obtained
\begin{subequations}
    \label{eq:estimation:recursive_bayesian:ss_model}
    \begin{align}
        \vec{x}[k]
            & = \boldsymbol{F} \vec{x}[k-1] + \vec{w}[k], \\
        z[k]
            & = \boldsymbol{G} \vec{x}[k] + \nu[k],
    \end{align}
\end{subequations}
where $\boldsymbol{F} = \boldsymbol{I}_{2N+1}$ is the $2N+1 \times 2N+1$ identity matrix and $\boldsymbol{G}$ 
is the $1 \times 2N+1$ dimensional observation matrix with the $i$\textsuperscript{th} component $g_i$ defined as
\begin{equation*}
g_i = 
\begin{cases}
 1, & i = 1, \\
 \sin(2\pi f_{i-1}t_k), & 1 < i \le N_{\text{KF}} + 1, \\
 \cos(2 \pi f_{i - N_{\text{KF}} - 1} t_k), & N_{\text{KF}} + 1 < i \le 2 N_{\text{KF}} + 1.  
\end{cases}
\end{equation*}

The linear model in Eq.~\eqref{eq:estimation:recursive_bayesian:ss_model} can then be used in a Kalman 
filter~\cite{Sarkka2013} to obtain recursive estimates of the Fourier coefficients $\vec{x}[k]$ at each time $t_k$. 
Finally, a point estimate of the breathing frequency is obtained in the same way as for the spectrum based method, that 
is, by selecting the frequency with highest magnitude such that
\begin{equation}
    \hat{f}_{\text{KF}}
        = \argmax_{n} \sqrt{a_n^2 + b_n^2}, \qquad n > 0.
    \label{eq:estimation:recursive_bayesian:point_estimate}
\end{equation}
Note that this method has all the advantages of recursive implementations including decreased computational and memory 
requirements. In addition to these, an important advantage of this method compared to the DFT-based approach is its 
capability of working with unevenly sampled data.

\subsection{Model-based Estimation}\label{sec:estimation:model}
In principle, the model of the RSS measurements in Eq.~\eqref{eq:R-linear-scale-cos-sine} and 
Eq.~\eqref{eq:R-log-scale-cos-sine} 
could be exploited to improve the perturbation frequency estimation or to relax the requirements of the DFT based 
(both batch or recursive) estimators. However, since the small perturbations depend on many different parameters such as
the breathing direction with respect to link-line, breathing amplitude, initial position of the person, their geometry, 
 and electrical properties of their clothes this is non-trivial. Instead, non-parametric statistical models can be used 
 for
 capturing these effects. This approach has recently been used in~\cite{Hostettler2017} by modeling the underlying 
 signal 
 as a quasi-periodic Gaussian process. Here, we just summarize 
the main results of this model-based method, and the reader is referred to~\cite{Hostettler2017} for related 
discussions and a detailed derivation.

The low-pass filtered RSS $z[k]$ (see Fig.~\ref{fig:preprocessing}) can be modeled as a quasi-periodic Gaussian 
process~\cite{Rasmussen2006,Solin2014}, such that
\begin{subequations}
    \label{eq:estimation:model:gp_model}
    \begin{align}
        z[k]
            & = g[k] + \nu[k], \\
        g(t)
            & \sim \GP(0, K(\tau)),
    \end{align}
\end{subequations}
where $g[k] \triangleq g(t_k)$ is the $k$\textsuperscript{th} sample acquired at time $t_k$, and $\GP(m(t), K(t, t'))$ 
denotes a Gaussian process prior with mean function $m(t)$ and covariance kernel $K(t, t') = K(t-t')$, and $\tau = t - 
t'$~\cite{Rasmussen2006}. For a temporal Gaussian process with the canonical periodic covariance function given by
\begin{equation}
    K(\tau)
        = \sigma_K^2 \exp\left(-\frac{2 \sin^2\left(\frac{2\pi f \tau}{2}\right)}{\ell^2}\right),
    \label{eq:estimation:model:covariance_kernel}
\end{equation}
%
it can be shown that the following equivalent discrete-time state-space 
formulation
\begin{subequations}
    \label{eq:estimation:model:ss_gp_model}
    \begin{align}
        u_0[k] &= u_0[k-1] + w_0[k], \\
        \vec{u}_n[k] & = \boldsymbol{F}_n \vec{u}_n[k-1] + \vec{w}_n[k-1], \\
        g[k] & = u_0[k] + \sum_{n = 1}^\infty \boldsymbol{H}_n \vec{u}_n[k]
        \label{eq:estimation:model:ss_gp_model:decomposition}
    \end{align}
\end{subequations}
can be obtained~\cite{Solin2014,Hostettler2017}. In Eq.~\eqref{eq:estimation:model:covariance_kernel} $\sigma_K^2$ 
(variance), $\ell$ (length scale), and $f$ (perturbation frequency) are hyperparameters. In 
Eq.~\eqref{eq:estimation:model:ss_gp_model}, $u_0[k]$ is the DC component, 
$\vec{u}_n[k]$ (for $n > 0$) is a $2 \times 1$ vector containing the instantaneous value of the $n^\text{th}$ harmonic 
and its derivative, ${w}_{0}[k] \sim \N(0, {C}_{w_0})$ and $w_n[k] \sim \N(0, \boldsymbol{C}_{w_n})$ are the 
corresponding process noises, and
\begin{subequations}\label{eq:estimation:gp-definitions}
    \begin{align}
        \boldsymbol{F}_n
            & = \begin{bmatrix}
                    \cos(2\pi f_n \delta_t) & -\sin(2\pi f_n \delta_t) \\
                    \sin(2\pi f_n \delta_t) &  \cos(2\pi f_n \delta_t)
        \end{bmatrix}, \\
        \boldsymbol{H}_n
            & = \begin{bmatrix} 1 & 0 \end{bmatrix}, \\
        \boldsymbol{C}_{w_n}
            & = 4 \delta_t \sigma_K^2 \exp(-\ell^{-2}) \mathcal{I}_n(\ell^{-2}) \boldsymbol{I}_2, \\
        {C}_{w_0} &= 2 \delta_t \sigma_K^2 \exp(-\ell^{-2}) \mathcal{I}_n(\ell^{-2}),
    \end{align}
\end{subequations}
where $\mathcal{I}_n(\cdot)$ is the $n$\textsuperscript{th} order modified Bessel function of the first kind, and 
$\delta_t = t_k - t_{k-1}$. The initial states are given by $\gamma_{0}[0] \sim \N(0, P_{0,\gamma_0})$ and 
$\boldsymbol{u}_n[0] \sim \N(0, \boldsymbol{P}_{0,\vec{u}_n})$.

Additionally, the logarithm of the breathing frequency $s[k] = \log(f[k])$ is modeled as a geometric Brownian motion 
which yields~\cite{Oksendal2010}
\begin{equation}
    s[k] = s[k-1] - \frac{1}{2} S_f^2 \delta_t + w_{s}[k],
    \label{eq:estimation:model:frequency_model}
\end{equation}
where $S_f$ is the spectral density of the underlying white noise process and $w_{s}[k] \sim \N(0, S_f \delta_t)$.By 
introducing a time-varying frequency as in Eq.~\eqref{eq:estimation:model:frequency_model}, this assumption is violated 
for the covariance kernel in Eq.~\eqref{eq:estimation:model:covariance_kernel}. However, since the breathing rate 
varies relatively slowly, the process is considered locally stationary.

Finally, combining Equations~\eqref{eq:estimation:model:gp_model},~\eqref{eq:estimation:model:ss_gp_model}, 
and~\eqref{eq:estimation:model:frequency_model}, and truncating the 
series in Eq.~\eqref{eq:estimation:model:ss_gp_model:decomposition} at some upper bound $N_{\text{GP}}$, the following 
nonlinear 
state-space model is obtained
\begin{subequations}
    \label{eq:estimation:model:complete_gp_model}
    \begin{align}
        \begin{bmatrix}
            s[k] \\
            u_0[k] \\
            \vec{u}_1[k] \\
            \vdots \\
            \vec{u}_{N_\text{GP}}[k]
        \end{bmatrix}
            & = \begin{bmatrix}
                    s[k-1] - \frac{1}{2} S_f^2 \delta_t \\
                    u_0[k-1] \\
                    \boldsymbol{F}_1 \vec{u}_1[k-1] \\
                    \vdots \\
                    \boldsymbol{F}_{N_\text{GP}} \vec{u}_{N_\text{GP}}[k-1]
                \end{bmatrix} +
                \begin{bmatrix}
                    w_s[k] \\
                    w_{0}[k] \\
                    \vec{w}_{1}[k] \\
                    \vdots \\
                    \vec{w}_{N_\text{GP}}[k]
                \end{bmatrix}, \\
        z[k]
            & = u_0[k] + \sum\limits_{n = 1}^{N_{\text{GP}}} \boldsymbol{H}_n \vec{u}_n[k] + \nu[k],
    \end{align}
\end{subequations}
where the dependence of the matrices $\boldsymbol{F}_n$ on $s[k]$ is implicit. The 
model Eq.~\eqref{eq:estimation:model:complete_gp_model} can now readily be used in a (nonlinear) Kalman filter such as 
the unscented Kalman filter or extended Kalman filter~\cite{Sarkka2013}. In this paper, we use Rao--Blackwellized 
unscented Kalman Filter presented earlier in~\cite{Hostettler2017}. Finally, the frequency estimate is given by
\begin{equation}
\hat{f}_\text{GP} = \exp(\hat{s}),
\end{equation}
where $\hat{s}$ is the first component of the latest state estimate.

\begin{figure*}[t]
	\begin{center}
		\begin{tabular}{cccc}
			\subfloat[RSS on channel 3]{\includegraphics[,width=4.5cm,height=4cm]{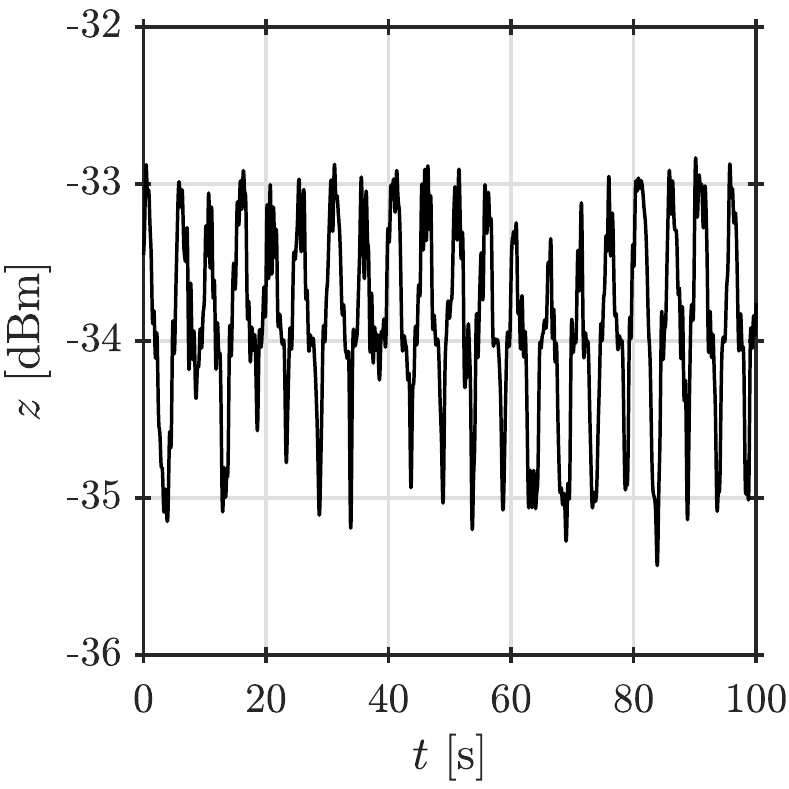}\label{fig:rss_1}}
			\subfloat[GP state 
			estimates]{\includegraphics[,width=4.5cm,height=4cm]{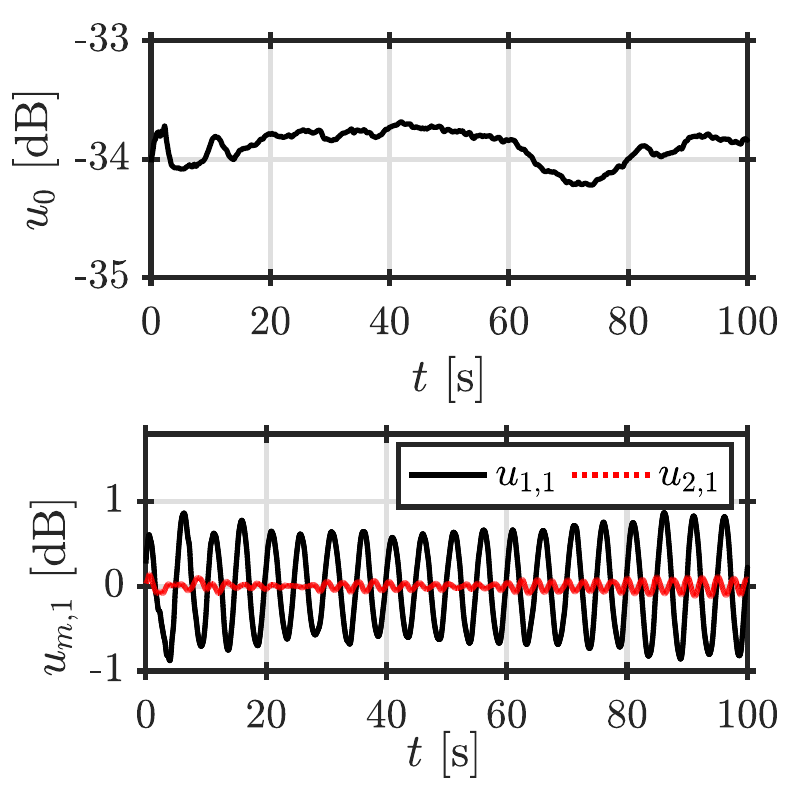}\label{fig:gp_coefficients_1}}
			\subfloat[Spectrograms]{\includegraphics[,width=4.5cm,height=4cm]{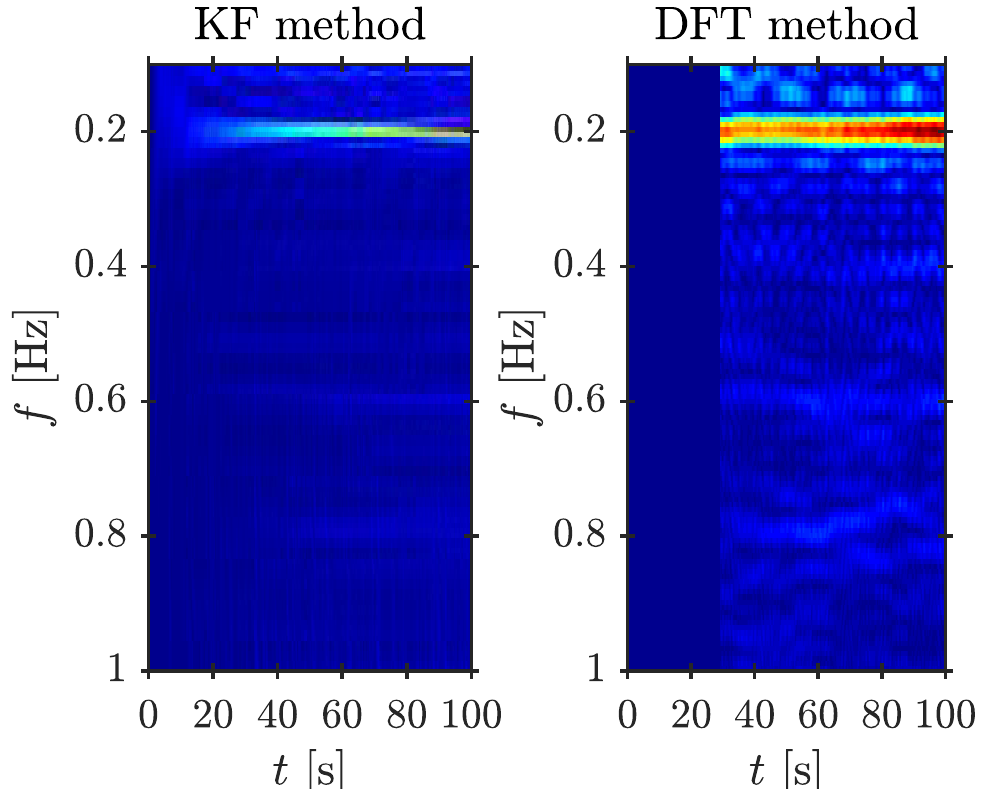}\label{fig:spectrogram_1}}
			\subfloat[Breathing rate 
			estimates]{\includegraphics[width=4.5cm,height=4cm]{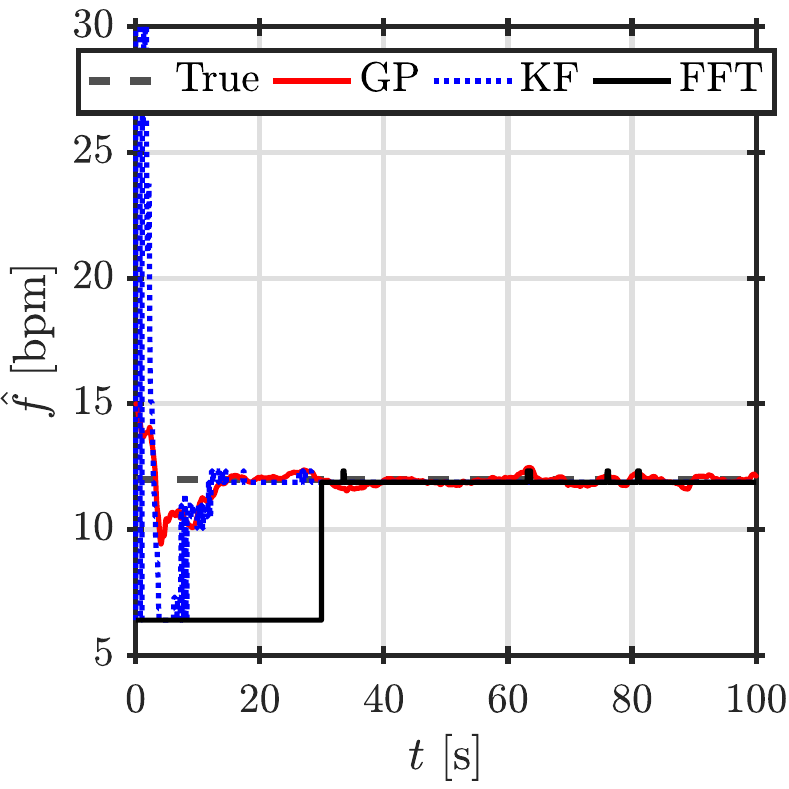}\label{fig:breathing_estimate_1}}
			\\
			\subfloat[RSS on channel 10]{\includegraphics[width=4.5cm,height=4cm]{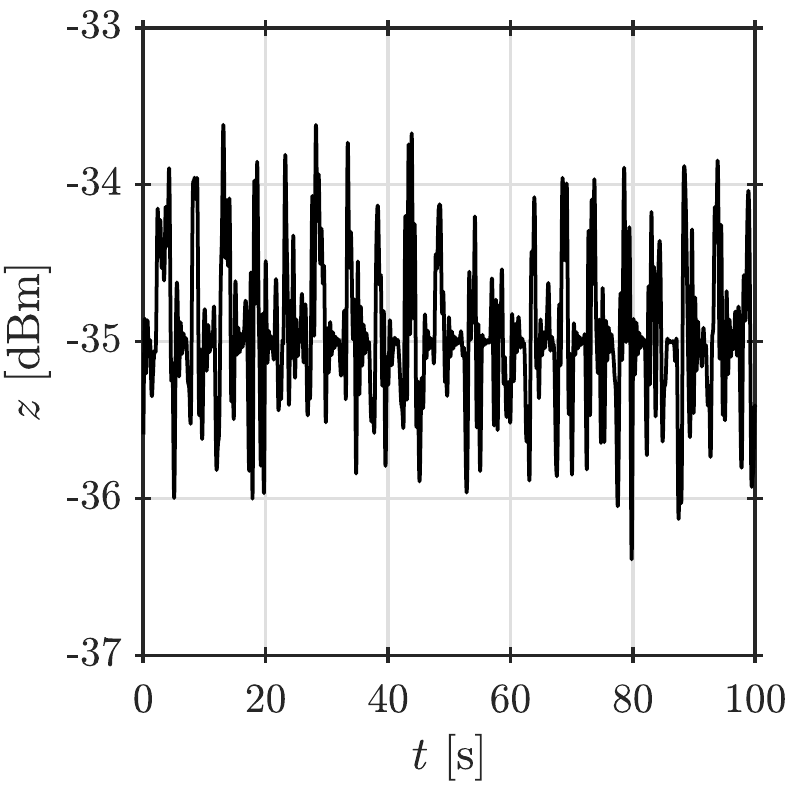}\label{fig:rss_2}}
			\subfloat[GP state 
			estimates]{\includegraphics[width=4.5cm,height=4cm]{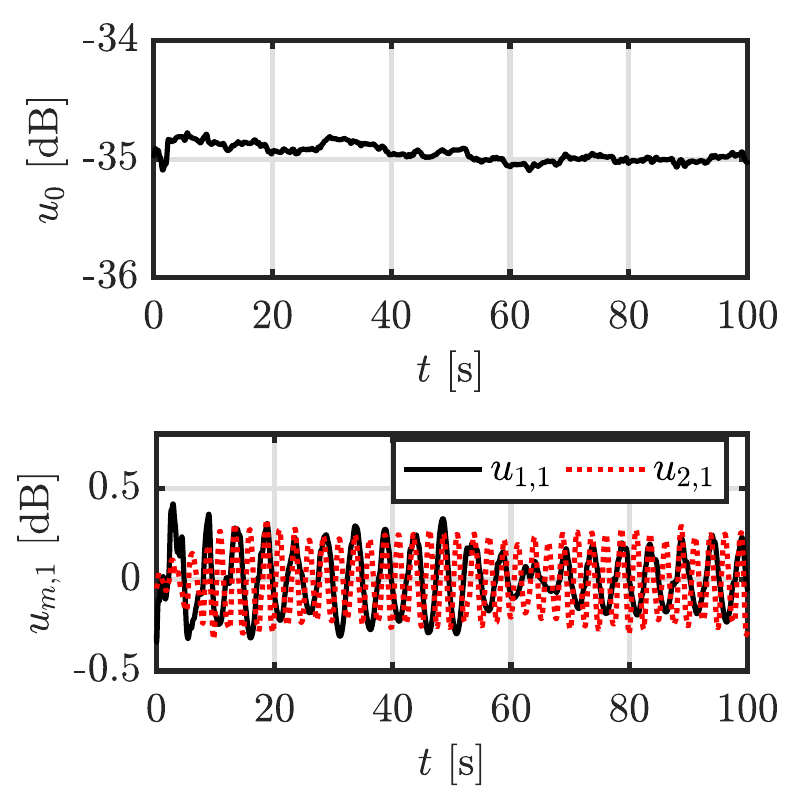}\label{fig:gp_coefficients_2}}
			\subfloat[Spectrograms]{\includegraphics[width=4.5cm,height=4cm]{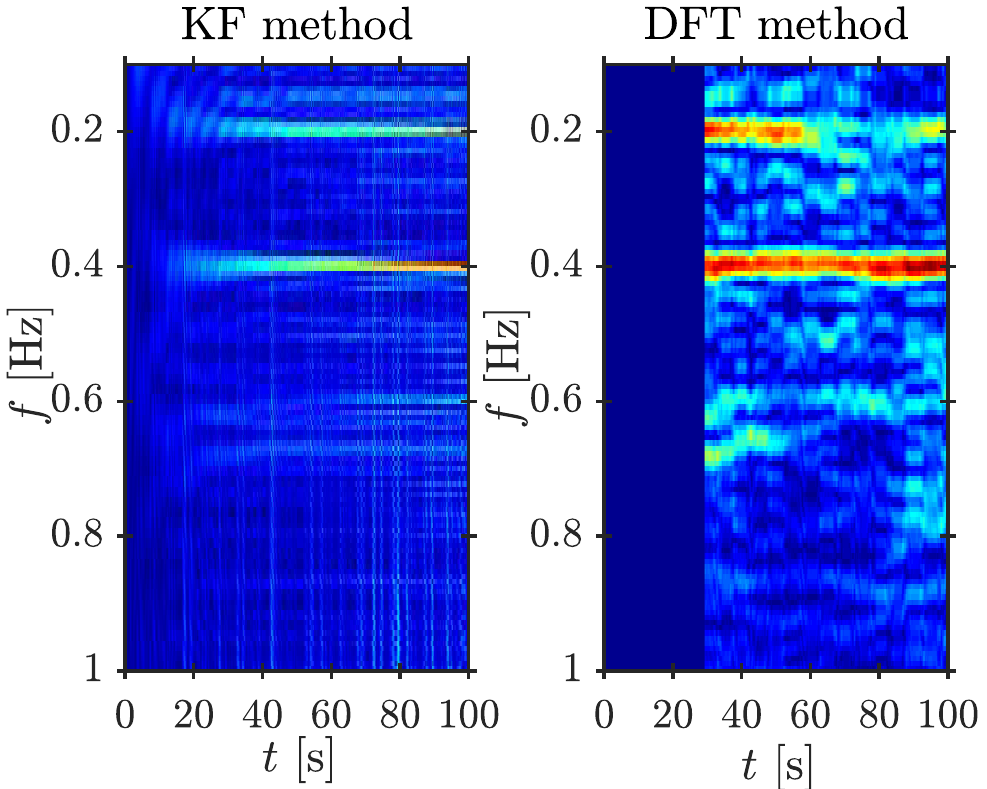}\label{fig:spectrogram_2}}
			\subfloat[Breathing rate 
			estimates]{\includegraphics[width=4.5cm,height=4cm]{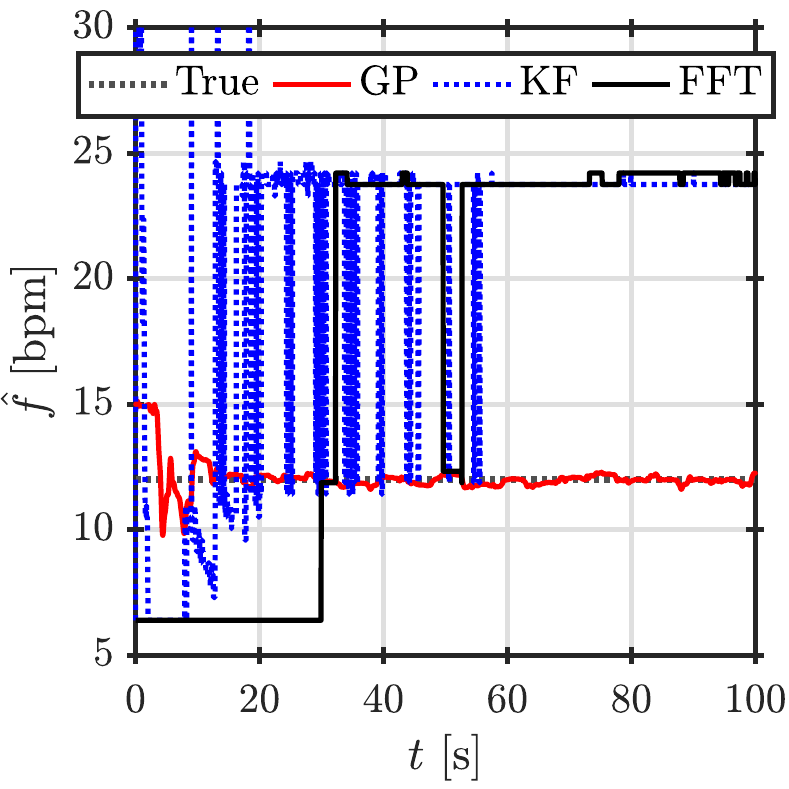}\label{fig:breathing_estimate_2}}
		\end{tabular}
		\caption{Bed experiment: breathing monitoring on two different frequency channels using the presented methods. 
		In (a) and (e), the breathing affected RSS signal. In (b) and (f), state estimates of the GP method 
		$(N_\text{GP}=2)$. Spectrograms of the spectral estimation techniques illustrated in (c) and (g). The breathing 
		rate estimates on the two channels are shown in (d) and (h). 
		}
	\label{fig:breathing_estimation}
	\end{center}
\end{figure*}

\begin{table}
	\caption{Evaluation Parameters} 
	\renewcommand{\arraystretch}{1.0}
	\centering 
	\begin{tabular}{| l | c | c | p{3.2cm}|}
		\hline
		\bfseries Symbol &\bfseries Value &\bfseries Appearance &\bfseries Explanation  \\ \hline
		$f$ & Varying  &  Sec.~\ref{sec:experiment-setup} & Breathing frequency in bpm\\ \hline
		$f_s$ & Varying  &  Sec.~\ref{sec:experiment-setup} & Sampling frequency in Hz\\ \hline
		$N_\text{DFT}$ & $2048$  & Eq.~\eqref{eq:estimation:windowed:psd} & Number of FFT points\\ \hline
		$N_w$ & $30$ s data & Eq.~\eqref{eq:estimation:windowed:data}& FFT window length  \\ \hline
		$N_o$& $N_w - 1$ & Eq.~\eqref{eq:estimation:windowed:data} & FFT window overlap \\ \hline
		$N_\text{KF}$& $75$ & Eq.~\eqref{eq:estimation:recursive_bayesian:fourier_series_expansion} & Number of 
		    frequency bins in KF \\ \hline
		$\boldsymbol{C}_w$ & $0.01 \boldsymbol{I}_{2 N + 1}$& Eq.~\eqref{eq:estimation:recursive_bayesian:ss_model} & 
			Process noise covariance \\ \hline
		$\boldsymbol{m}_{0, \text{KF}}$ & $\begin{bmatrix}z[0]\\ 0\\ \vdots\\ 0 \end{bmatrix}$ & 
			Eq.~\eqref{eq:estimation:recursive_bayesian:ss_model} & Initial estimate of KF \\ \hline 
		$\boldsymbol{P}_{0, \text{KF}}$ & $\boldsymbol{I}_{2 N + 1}$& 
			Eq.~\eqref{eq:estimation:recursive_bayesian:ss_model} & Covariance of the initial estimate \\ \hline
		$N_\text{GP}$& $2$ & Eq.~\eqref{eq:estimation:recursive_bayesian:fourier_series_expansion} & GP Truncation 
		Order \\ \hline
		$\sigma_K^2$ & $0.01$& Eq.~\eqref{eq:estimation:model:covariance_kernel} & Covariance kernel variance\\ \hline
		$\ell$ & $0.9$& Eq.~\eqref{eq:estimation:model:covariance_kernel} & Covariance kernel length scale \\ 
			\hline
		$S_f$ & $10 ^{-4}$& Eq.~\eqref{eq:estimation:model:frequency_model} & White noise process PSD \\ \hline
		$\delta_t$& $1/f_s$& Eq.~\eqref{eq:estimation:gp-definitions} & Time difference between samples in seconds\\ 
		\hline
		$P_{0,u_0}$ & $\sqrt{0.1}\boldsymbol{I}_2$& Eq.~\eqref{eq:estimation:gp-definitions} & Covariance of
			the initial state estimate components corresponding to DC terms \\ \hline
		$\boldsymbol{P}_{0,\boldsymbol{u}_n}$ & $\frac{1}{2^n n!}\boldsymbol{I}_2$& 
		Eq.~\eqref{eq:estimation:gp-definitions} 
			& Covariance of the initial state estimate components corresponding to non-DC terms \\ \hline
		$\boldsymbol{m}_{0, \text{GP}}$ & 
			$\begin{bmatrix}\ln\left( {15}/{60}\right) \\ z[0]\\ 0\\ \vdots\\ 0 \end{bmatrix}$ & 
			Eq.~\eqref{eq:estimation:model:complete_gp_model} & Initial estimate of GP \\ \hline
		$\sigma_\nu^2$ & $1$& Various & Measurement noise variance for KF and GP models \\ \hline 
	\end{tabular}
	\label{table:parameters} 
\end{table}

\section{Experimental Results}\label{sec:experiments}

In this section, the developments of the paper are evaluated using experimental data. In the following, we present the 
experimental setup and overview of the experiments before introducing the evaluation metrics. Then, the results are 
given.

\subsection{Experimental Setup and Experiments}\label{sec:experiment-setup}
The experiments are conducted using the nodes with the hardware and software 
platform described in~\cite{Yigitler2014}. A TX node is programmed to transmit 
packets over $16$ frequency channels at the $2.4$ GHz ISM band\footnote{The developments presented thus far consider a 
stream of RSS measurements with constant wavelength. In this section, the acquired measurements from different 
frequency channels are considered as different measurements with different wavelengths.}. After each transmission, the 
frequency channel of 
communication is changed sequentially to cover the $80$ MHz spectrum. The RX nodes are programmed to listen for ongoing 
transmissions. Upon reception, the packets are timestamped at the start of frame delimiter with a resolution of $1/32$ 
microseconds, and the received frames are stored to a non-volatile memory. 

The first experiment aims at evaluating the accuracy of the estimators and the experimental procedure follows
that of experiment no.\ 1 in~\cite{Kaltiokallio2014}. The RSS measurements are acquired by a
single RX node, which is $2 \text{ m}$ away form the TX node. The 
transmission interval is set to $2 \text{ milliseconds}$. Thus, $f_s = 31.25 \text{ Hz}$ for each frequency 
channel, which is considerably higher than the breathing frequencies of interest. During the experiments the person is 
lying on a bed, where his chest is approximately $25 \text{ cm}$  away from the link-line while breathing at a constant 
rate set by a metronome. The person breaths at $5$ different rates: $12, 14, 16, 18 \text{ and } 20$ breaths-per-minute 
(bpm). In total $80$ time series are recorded and used for evaluation purposes. We refer to this experiment as 
\emph{bed experiment}. 

The second experiment aims at evaluating how the SNR impacts estimation performance. The TX node emits frames 
every $1.92 \text{ milliseconds}$, and $11$ RX nodes acquire the RSS variation when the person is standing in four 
different positions as shown in Fig.~\ref{fig:experiment_map}. At each position, the person is standing still while 
breathing at a constant rate of $12$ bpm. The setup is also used for acquiring RSS measurements when the environment is 
empty. In total $176$ time series are recorded for each position so that the evaluation is based on $704$ data series. 
We refer to this experiment as \emph{room experiment}. 

The acquired RSS data is preprocessed as shown in Fig.~\ref{fig:preprocessing}. The bandpass filtered data $y[k]$ is 
obtained as output of two processing stages: first the mean is calculated and removed from measurement 
$\mathcal{P}[k]$, and then the result is low-pass filtered. The low-pass filtered data $z[k]$ is obtained by just 
low-pass filtering $\mathcal{P}[k]$. The used low pass filter is a $5$\textsuperscript{th} order elliptic filter that 
has passband frequency of $2 \text{ Hz}$ and stop frequency of $3 \text{ Hz}$, $0.05 \text{ dB}$ maximum ripple in the 
pass band, and $40 \text{ dB}$ stop band attenuation. 

\subsection{Evaluation Methodology}

During the evaluation, we refer to the output of the batch 
estimator summarized 
in Sec.~\ref{sec:estimation:windowed} as DFT estimate, the recursive Bayesian estimator presented in 
Sec.~\ref{sec:estimation:recursive_bayesian} as Kalman Filter 
(KF) estimate, and finally, the model-based estimator presented in Sec.~\ref{sec:estimation:model} output is referred 
to as Gaussian Process (GP) estimate. These methods are implemented using the parameters tabulated in 
Table~\ref{table:parameters}.

The evaluation in this section is based on mean absolute error (MAE) calculations. For the breathing frequency 
estimation, the MAE in bpm is defined as
\begin{equation}
\varepsilon_f \triangleq \frac{60}{K}\sum\limits_{k=1}^{K} |\hat{f}[k] - f|,
\end{equation}  
where $f$ is the true rate, and $\hat{f}[k]$ is the $k$\textsuperscript{th} frequency estimate out of $K$ total
estimates. This metric fails to provide a measure of dispersion in the estimates. For this 
purpose, we use the ratio of estimates within $1$ bpm neighborhood of the true frequency $f$
\begin{equation}
\varepsilon_{\%} \triangleq \frac{\text{\# of $\hat{f}$ in $1$ bpm neighborhood of $f$}}
{\text{\# of estimates}}  \cdot 100.
\end{equation}
In order to quantify the convergence speed of the methods, we calculate $\varepsilon_f$ for the data in the first 
$30 \text{ seconds}$ and for the data afterwards separately. We refer to the former as $\varepsilon_f(t \le 30 \text{ 
s})$ and the latter as $\varepsilon_f(t > 30 \text{ s})$. When the estimates have some outliers, e.g. they converge to 
the second harmonic frequency, we also calculate $\varepsilon_f$ by excluding those outliers, and refer the result with 
\emph{$\varepsilon_f$ w/o outliers}.

In high SNR conditions, the $\varepsilon_f$ performance of all the estimators are similar. In this case, the MAE of the 
estimated signal and the estimator input can be used for evaluation, since the estimator outputs also imply a signal in 
a specific form. Let us denote the model output of any estimator\footnote{Although the 
state space models in Sec.~\ref{sec:estimation} are different, all of these yield the same output form as in 
Eq.~\eqref{eq:R-log-scale-cos-sine} only with different number of harmonics.} as $\hat{\mathcal{R}}$. Then, we define 
the modeling MAE as 
\begin{equation}
\varepsilon_z = \frac{1}{K}\sum\limits_{k=1}^K|z[k] - \hat{\mathcal{R}}[k]|,
\end{equation}   
where for DFT-based method we add the mean value subtracted in the preprocessing stage.

The room experiment is used for evaluating the performance of the estimators under varying SNR conditions. The SNR of 
the signal is estimated using the PSD estimate in Eq.~\eqref{eq:estimation:windowed:psd}, using the actual breathing 
frequency. Let set $\mathpzc{L}(f)$ contain the indices of the bins that are in a neighborhood of the harmonics of the 
true breathing frequency, excluding the DC term, and $\mathpzc{S}$ denote the bins within interval $0.1$ and $3$ Hz, 
which define the frequency range we are interested in. Then, an SNR estimate is given by
\begin{equation}
\hat{\varrho} \triangleq 10 \log_{10}\left\{{\sum\limits_{l \in \mathpzc{L}(f)}\boldsymbol{S}[l]} \bigg/
{\sum\limits_{l \in \mathpzc{S}/\mathpzc{L}(f)}\boldsymbol{S}[l]} \right\},
\end{equation}
where the sets are disjoint. In the following, we use only the first two harmonics when forming the set 
$\mathpzc{L}(f)$, 
and all other spectral bins contribute to the noise power. 

\subsection{Results}
\subsubsection{Bed experiment}

In Fig.~\ref{fig:breathing_estimation}, breathing estimation is illustrated on two different frequency channels when 
the 
person is breathing at a constant rate of $0.2 \text{ Hz}$, that is, $12 \text{ bpm}$. The signal shown in 
Fig.~\ref{fig:rss_1} contains a strong first harmonic at the breathing frequency and all methods are capable of 
estimating the true frequency correctly as illustrated in Fig.~\ref{fig:breathing_estimate_1}. The recursive GP and KF 
methods converge to the true frequency in approximately $15 \text{ s}$, whereas the DFT method requires $30 \text{ s}$ 
because of the time window used to calculate the DFT. The breathing induced changes are not as evident for the signal 
shown in Fig.~\ref{fig:rss_2} since it contains higher order harmonics as proposed by the model and as illustrated in 
Fig.~\ref{fig:spectrogram_2}. The KF and DFT methods estimate the breathing frequency using the peak of the spectrum, 
resulting in an incorrect estimate of $f \approx 24 \text{ bpm}$ which corresponds to the second harmonic. The higher 
order harmonics are taken into account in the GP-based estimator when truncation order is higher than $1$, 
$N_{\text{GP}} > 1$. As a result, the method can correctly estimate the 
true breathing frequency as illustrated in Fig.~\ref{fig:breathing_estimate_2}. The state estimates of the GP are 
illustrated in Fig. \ref{fig:gp_coefficients_2} and clearly, the method is able to track the DC-component and the 
harmonics accurately resulting in an improvement with respect to the spectral estimation techniques. For clarity, the 
second component of $\boldsymbol{u}_j$ is omitted from Figs.~\ref{fig:gp_coefficients_1} and 
\ref{fig:gp_coefficients_2}.

\begin{table}
	\caption{Results of the bed experiment} 
	\renewcommand{\arraystretch}{1.2}
	\centering 
	\begin{tabular}{| l | c | c | c |}
		\hline
		& GP & KF & DFT \\ \hline
		$\varepsilon_\%$ [\%] & 97.50 & 88.75 & 87.50 \\ \hline
		$\varepsilon_f(t \le 30 \text{ s})$ [bpm] & 1.06 & 5.10 & - \\ \hline
		$\varepsilon_f(t > 30 \text{ s})$ [bpm] & 0.25 & 1.44 & 1.41 \\ \hline
		$\varepsilon_f$ w/o outliers [bpm] & 0.15 & 0.25 & 0.26 \\ \hline
		$\varepsilon_z$ [dB] & 0.16 & 0.32 & 0.28 \\ \hline
	\end{tabular}
	\label{table:bed_experiment} 
\end{table}

The measurement setup of the bed experiment is a realization of the measurement setup evaluated in 
Fig.~\ref{fig:harmonic_energy}. Thus, small displacements of the person, as can be observed in 
Fig.~\ref{fig:harmonic_energy}, causes drastic changes in the SNR. Furthermore, different frequency channel 
measurements may have different behavior as the spread of the dots in the figure imply. For the acquired $80$ signals, 
there are $8$ signals having higher energy in the second harmonic. According to the evaluation in 
Fig.~\ref{fig:harmonic_energy}, a particular $y$ value (between $0.24$ and $0.26$ m) can yield such measurements. 
Therefore, the model correctly resembles this important scenario. 

Performance of the estimators is summarized in Table~\ref{table:bed_experiment} and on average, the spectral estimation
techniques yield comparative accuracy while the GP-based estimator outperforms them. The ratio of valid estimates is
summarized by $\varepsilon_\%$ and the $10\%$ difference in favor of the GP method originates from the fact that the 
second
harmonic has the highest amplitude in $8$ out of $80$ signals resulting incorrect estimates with the spectral
techniques. The recursive GP and KF methods typically converge in the first $30 \text{ s}$ and $75 \%$ of the estimates
converge to within $1 \text{ bpm}$ of the true rate in $15.8 \text{ s}$ with GP and in  $17.6 \text{ s}$ with KF. 
However,
the GP attains a significantly lower $\varepsilon_f(t \le 30 \text{ s})$ compared to KF, since the KF errors are 
typically
very large due to the jumpy behavior as observed in Fig.~\ref{fig:breathing_estimate_1}. After the transient period 
$(0 - 30 \text{ s})$, the GP achieves a lower MAE than the spectral estimation techniques as given by 
$\varepsilon_f(t > 30 \text{ s})$. However, these results are severely affected by the experiments that resulted in
incorrect estimates due to 
measurements not showing the first harmonic, but the second one. Neglecting these outliers, one can observe that all 
methods
yield comparative accuracy as given in the fifth row of Table~\ref{table:bed_experiment}. The steady-state accuracy of 
the
spectral methods is mainly affected by the frequency bin size, whereas the GP accuracy could be improved by selecting 
the
spectral density $S_f$ smaller. However, this would also decrease responsiveness of the filter to possible breathing 
rate
changes. Lastly, $\varepsilon_z$ is given in the last row of Table~\ref{table:bed_experiment}. Clearly, the GP model
estimates correspond more closely to the measured RSS since the higher order harmonics are taken into account. 

The development in Sec.~\ref{sec:models} concludes that the RSS is composed of more than one harmonics. However, the 
relative importance of higher order harmonics depend on several factors, which include actual breathing function (in 
this regard it is evident that natural breathing is not a sinusoid) and effect of quantization in typical RSS 
measurement systems~\cite{Yigitler2017}. In order to quantify the importance of higher order harmonics, one may 
investigate the estimated energy in harmonics for all $80$ time series. 

\begin{table}
	\caption{Truncation order} 
	\renewcommand{\arraystretch}{1.2}
	\centering 
	\begin{tabular}{| l | c | c | c | c |}
		\hline
		Truncation order $N_\text{GP}$ & 1 & 2 & 3 & 4 \\ \hline
		$\varepsilon_\%$ [\%] & 97.50 & 97.50 & 97.50 & 97.50 \\ \hline
		$\varepsilon_f(t > 30 \text{ s})$ [bpm] & 0.28 & 0.29 & 0.25 & 0.25 \\ \hline
		$\varepsilon_z$ [dB] & 0.19 & 0.17 & 0.16 & 0.16 \\ \hline
	\end{tabular}
	\label{table:truncation_order} 
\end{table}

Since the model in Eq.~\eqref{eq:estimation:model:ss_gp_model:decomposition} is composed of Fourier series coefficients 
(and their derivatives), the relative energy in the $m$\textsuperscript{th} harmonic can be defined as 
\begin{equation*}
\mathcal{E}_{(m)} = {u_{m,1}^2}\Big/{\sum\limits_{n=1}^{4}{u_{k,1}^2}},
\end{equation*}
where $u_{m,1}$ denotes the first component of $\vec{u}_m$.

Averaging $\mathcal{E}_{(m)}$ across the $80$ experiments results to $\mathcal{E}_{(1)} = 86.62\%, \mathcal{E}_{(2)} = 
11.00\%, \mathcal{E}_{(3)} = 2.13\% \text{ and } \mathcal{E}_{(4)} = 0.25\%$, thus the first two harmonics 
contain approximately $98\%$ of the energy. This value is very close to the value predicted by the Carson's rule of 
thumb, and implies that the actual breathing signal is a smooth function, not containing any jumps. Typically 
$\mathcal{E}_{(1)} \approx 93\%$, but it can be as low as $36\%$ validating the 
importance of having an estimator that takes into account the higher harmonics. The accuracy of the GP with different 
truncation orders is given in Table \ref{table:truncation_order}. In terms of estimation accuracy, higher truncation 
order yields slightly better performance as indicated by $\varepsilon_f(t > 30 \text{ s})$. In addition, 
$\varepsilon_z$ is reduced with higher truncation order.

\subsubsection{Room experiment}

\begin{figure*}[t]
	\begin{center}
		\begin{tabular}{ccc}
			\subfloat[]{\includegraphics[width=6cm,height=4.81cm]{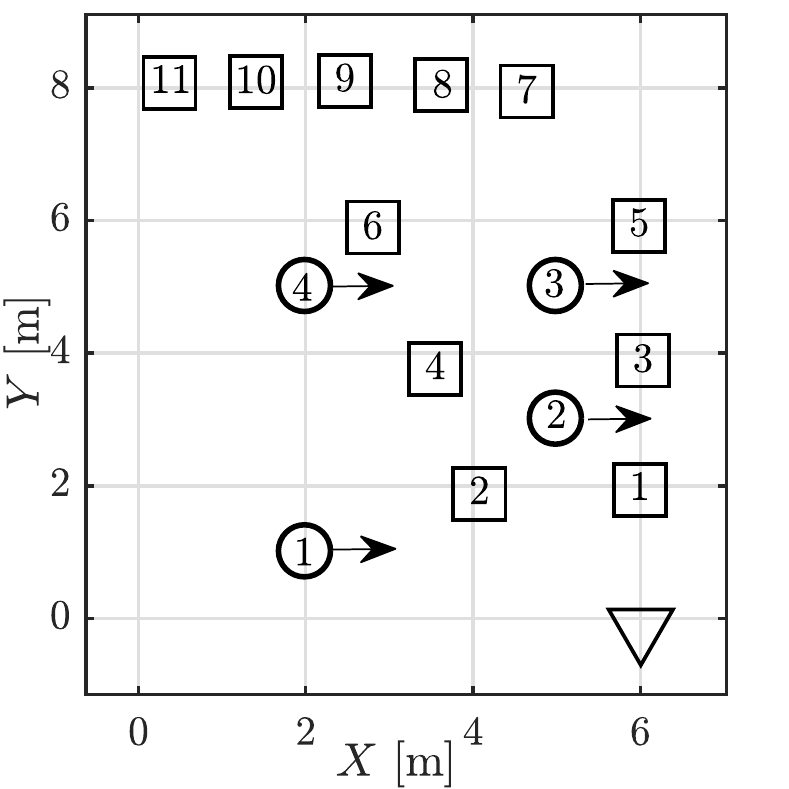}\label{fig:experiment_map}}
			\subfloat[]{\includegraphics[width=6cm,height=4.8cm]{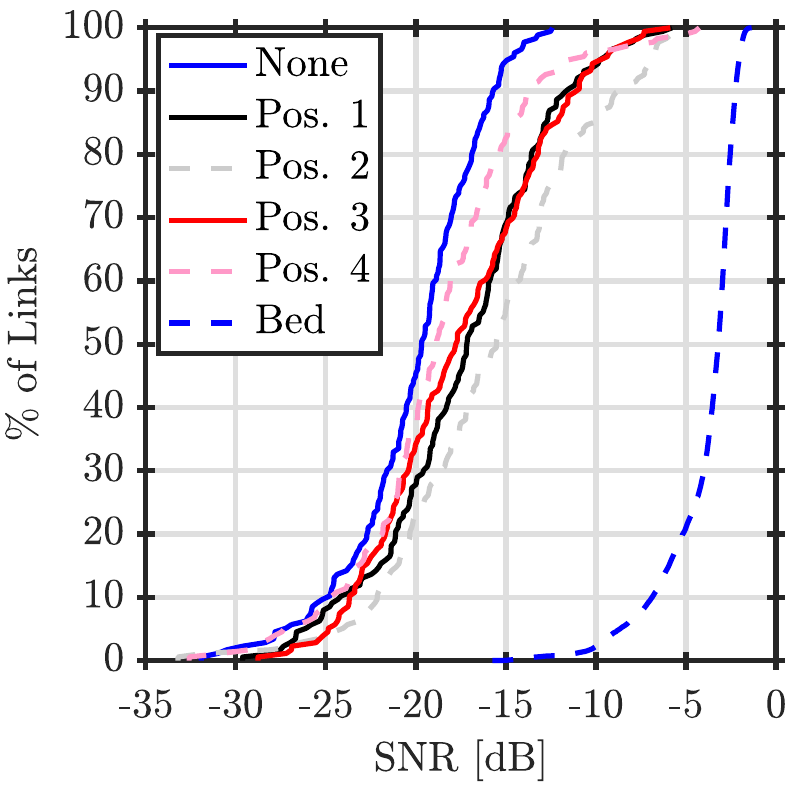}\label{fig:SNR_vs_position}}
			\subfloat[]{\includegraphics[width=6cm,height=4.8cm]{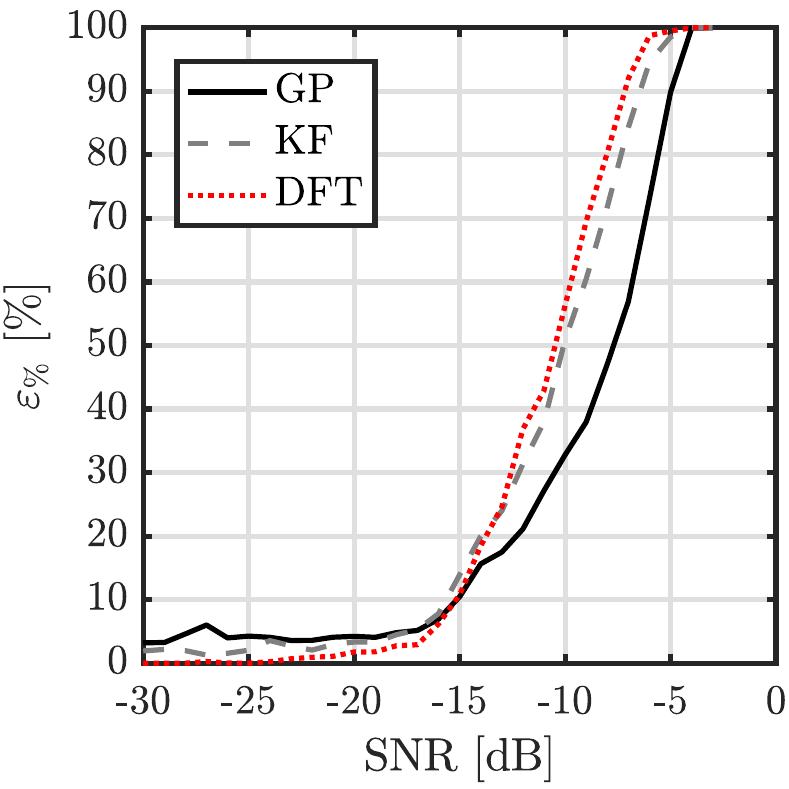}\label{fig:SNR_vs_estimator}}
		\end{tabular}
		\caption{In (a), layout of the room experiment, where $\bigcirc $ is the position number of the person while 
		facing to the direction shown with an arrow, 
		$\bigbox$ is the receiver node identifier between $1$ and $11$, and finally $\bigtriangledown$ is 
		the transmitter node. CDF of link SNRs in different positions, including empty room (None) and bed experiment 
		(Bed), are shown in (b). 
		In (c), the variation of $\varepsilon_\%$ with SNR for the different estimators is shown.
		}
		\label{fig:SNR}
	\end{center}
\end{figure*}

CDFs of link SNRs for the room experiment are illustrated in Fig.~\ref{fig:SNR_vs_position}. The maximum SNR value for
the empty room measurements is $-12.42 \text{ dB}$ and $7.4-25.0 \%$ of link SNRs exceeds this value when a breathing 
person is present. Thus, all positions clearly contain information regarding respiration rate of the person. 
However, SNR in the room experiments is typically much lower than in the bed experiment as shown in 
Fig.~\ref{fig:SNR_vs_position}.

\begin{table}
	\caption{ $\varepsilon_\%$ [\%] with DFT method in room experiment} 
	\renewcommand{\arraystretch}{1.2}
	\centering 
	\begin{tabular}{| c | c | c | c | c | c | c | c | c | c | c | c |}
		\hline
		& \multicolumn{11}{c|}{Receiver number} \\ \hline
		pos. & 1 & 2 & 3  & 4  & 5  & 6  & 7  & 8  & 9  & 10 & 11  \\ \hline
		1    & 6 & 0 & 13 & 0  & 6  & 19 & 19 & 19 & 25 & 19 & 6    \\ \hline
		2    & 0 & 6 & 75 & 44 & 13 & 31 & 6  & 6  & 38 & 6  & 13    \\ \hline
		3    & 0 & 0 & 6  & 6  & 25 & 13 & 6  & 25 & 6  & 6  & 25      \\ \hline
		4    & 0 & 0 & 0  & 6  & 19 & 0  & 13 & 0  & 6  & 0  & 31   \\ \hline
	\end{tabular}
	\label{table:room_experiment} 
\end{table}

The SNR of a link is defined by the signal energy  $\mathcal{E}_1$ given in Eq.~\eqref{eq:Energy-log-scale} for which 
the coefficients $\mathpzc{c}_m$ can be calculated using Eq.~\eqref{eq:R-log-scale-coefficients}. The coefficients 
$c_m$ are functions effective reflection coefficient $G$, defined in Eq.~\eqref{eq:G-definition}, and effective 
amplitude of the periodic movement  $\tilde{A}$ given in Eq.~\eqref{eq:A-tilde-phi}. 
The signal energy $\mathcal{E}_1$ increases as $G$ increases, and $G$ has its maximum when $\Gamma$ is at  its maximum  
on the link line. In the room experiment, when the person is on position~$1$, $G$ is in the interval $[0.02, \; 0.09]$ 
for nodes $1$ to $5$ ($1-5$), whereas for nodes $7 - 11$, $G \in [0.11, \; 0.23]$. As a result, the calculated signal 
energy $\mathcal{E}_1 = [0.02, \;0.16]$ for RX nodes $1-5$ and $\mathcal{E}_1 \in [0.19, \;0.23]$ for nodes $7-11$. 
With nodes $1-5$, the average SNR is $-18.22 \text{ dB}$ and $\varepsilon_\%=5.0\%$ whereas with RX nodes $7-11$, the 
average SNR is $-16.24 \text{ dB}$ and $\varepsilon_\%=17.5\%$. Thus, in position~$1$, it is expected that successful 
breathing monitoring is more likely with nodes $7-11$. The estimation results given in 
Table~\ref{table:room_experiment} are in accordance with this statement. 

The effective reflection coefficient $G$ is not the only parameter that affects the signal energy. In position~$2$, $G$ 
is in $[0.21, \; 0.33]$ for nodes $3-4$, whereas for nodes $7-8$, $G \in [0.70,\; 0.94]$. 
Respectively, the effective perturbation amplitude $\tilde{A} \in [0.30, \;0.54]$ for nodes $3-4$, it is in 
$[0.02,\;0.12]$ for $7-8$. Position~$2$ does not favor breathing monitoring using nodes $7-8$ despite that $G$ is 
three times larger than with nodes $3-4$. The reason for this short coming is 
that breathing causes very small changes in the RSS of nodes $7-8$ and $\mathcal{E}_1 \in [0.00, \;0.11]$, whereas 
for nodes $3-4$ the signal energy is $\mathcal{E}_1 \in [0.30, \;0.86]$ due to orientation of the person. 
Correspondingly, the average SNR of RS nodes $3-4$ is $-10.53 \text{ dB}$ and $\varepsilon_\%=59.4\%$, whereas for RX 
nodes $7-8$ the average SNR is $-15.88 \text{ dB}$ and $\varepsilon_\%=6.3\%$.  Again, the experimental results support 
the implications of the model.  

It is important to keep in mind that a slight change in position or orientation can have a significant impact on the 
signal energy of the links and therefore, spatial diversity or frequency channel diversity must be used to increase
likelihood of successful breathing monitoring. Results for the different receivers and positions are 
summarized in Table~\ref{table:room_experiment} and it can be concluded that successful breathing monitoring is very 
likely across a large area as long as the position and orientation of the person yield SNR higher than $-5$ dB as in 
the bed experiment.

In Fig.~\ref{fig:SNR_vs_estimator}, the variation of $\varepsilon_{\%}$ as a function of SNR using the different 
estimators is depicted. As shown, estimation accuracy improves with every estimator when the SNR increases and 
$\varepsilon_{\%} = 100\%$ with all methods when $\text{SNR} \geq -4 \text{ dB}$. Although for the bed experiment GP 
outperforms the other estimators, its performance is lower when the SNR is low. In this region, the second and higher 
order harmonics have lower power than the noise so that they are not as important as they are under high SNR 
conditions. Correspondingly, the batch DFT method and recursive spectral estimators outperform GP. This result suggests 
that for low SNR operating region batch DFT estimator is better whereas for high SNR conditions GP is better in terms 
of accuracy. It is also to be noted that GP has other advantages, most notably it relaxes data acquisition requirements 
by not requiring uniform sampling, operating with occasional packet losses, and better fusing the measurements from 
different communication channels~\cite{Hostettler2017}. Therefore, it has an utmost importance to investigate 
achievable performance, and to select an appropriate estimator.

\section{Conclusions}\label{sec:conclusion}
Breathing rate is an important vital sign of which continuous monitoring may help to identify serious problems before 
they actually occur. In this paper, a signal model for received signal strength based non-contact respiration rate 
monitoring systems using commodity wireless devices is presented. It is shown that the signal model for low-end 
communication devices has the same form as the one in high-end radar based solutions. The effect of linear movement has 
also been derived, and significance of physical parameters are shown and discussed. Real world measurements are used 
for evaluating the performances of three previously presented estimators, and the result is compared with the 
implications of the model. The model implications are in coherence with the findings, and show that respiration rate 
monitoring systems must be evaluated before deployment. The estimator must be selected according to expected 
signal-to-noise ratio of the measurements and the constraints imposed by the hardware and software implementations of 
the wireless nodes.

\bibliographystyle{IEEEtran}

\begin{thebibliography}{10}
	\providecommand{\url}[1]{#1}
	\csname url@samestyle\endcsname
	\providecommand{\newblock}{\relax}
	\providecommand{\bibinfo}[2]{#2}
	\providecommand{\BIBentrySTDinterwordspacing}{\spaceskip=0pt\relax}
	\providecommand{\BIBentryALTinterwordstretchfactor}{4}
	\providecommand{\BIBentryALTinterwordspacing}{\spaceskip=\fontdimen2\font plus
		\BIBentryALTinterwordstretchfactor\fontdimen3\font minus
		\fontdimen4\font\relax}
	\providecommand{\BIBforeignlanguage}[2]{{%
			\expandafter\ifx\csname l@#1\endcsname\relax
			\typeout{** WARNING: IEEEtran.bst: No hyphenation pattern has been}%
			\typeout{** loaded for the language `#1'. Using the pattern for}%
			\typeout{** the default language instead.}%
			\else
			\language=\csname l@#1\endcsname
			\fi
			#2}}
	\providecommand{\BIBdecl}{\relax}
	\BIBdecl
	
	\bibitem{Cretikos2008}
	M.~A. Cretikos, R.~Bellomo, K.~Hillman, J.~Chen, S.~Finfer, and A.~Flabouris,
	``Respiratory rate: the neglected vital sign,'' \emph{Medical Journal of
		Australia}, vol. 188, no.~11, p. 657, 2008.
	
	\bibitem{AL-Khalidi2011}
	F.~Q. AL-Khalidi, R.~Saatchi, D.~Burke, H.~Elphick, and S.~Tan, ``Respiration
	rate monitoring methods: A review,'' \emph{Pediatric Pulmonology}, vol.~46,
	no.~6, pp. 523--529, 2011.
	
	\bibitem{Folke2003}
	M.~Folke, L.~Cernerud, M.~Ekstr{\"o}m, and B.~H{\"o}k, ``Critical review of
	non-invasive respiratory monitoring in medical care,'' \emph{Medical and
		Biological Engineering and Computing}, vol.~41, no.~4, pp. 377--383, 2003.
	
	\bibitem{Greneker1996}
	E.~F. Greneker, ``Radar sensing of heartbeat and respiration at a distance with
	security applications,'' in \emph{Proceedings of Radar Sensor Technology II
		(SPIE 3066)}, vol. 3066.\hskip 1em plus 0.5em minus 0.4em\relax SPIE, 1997,
	pp. 22 -- 27.
	
	\bibitem{Staderini2002}
	E.~M. Staderini, ``{UWB} radars in medicine,'' \emph{IEEE Aerospace and
		Electronic Systems Magazine}, vol.~17, no.~1, pp. 13--18, Jan 2002.
	
	\bibitem{Adib2015}
	F.~Adib, H.~Mao, Z.~Kabelac, D.~Katabi, and R.~C. Miller, ``Smart homes that
	monitor breathing and heart rate,'' in \emph{Proceedings of the 33rd Annual
		ACM Conference on Human Factors in Computing Systems}, ser. CHI '15.\hskip
	1em plus 0.5em minus 0.4em\relax New York, NY, USA: ACM, 2015, pp. 837--846.
	
	\bibitem{Liu2014}
	X.~Liu, J.~Cao, S.~Tang, and J.~Wen, ``{Wi-Sleep}: Contactless sleep monitoring
	via {WiFi} signals,'' in \emph{IEEE Real-Time Systems Symposium 2014}, Dec
	2014, pp. 346--355.
	
	\bibitem{Kaltiokallio2014}
	O.~Kaltiokallio, H.~Yi\u{g}itler, R.~J\"{a}ntti, and N.~Patwari, ``Non-invasive
	respiration rate monitoring using a single {COTS TX-RX} pair,'' in
	\emph{Proceedings of the 13th International Symposium on Information
		Processing in Sensor Networks, IPSN-14}, April 2014, pp. 59--69.
	
	\bibitem{Patwari2014a}
	N.~Patwari, J.~Wilson, S.~Ananthanarayanan, S.~K. Kasera, and D.~R. Westenskow,
	``Monitoring breathing via signal strength in wireless networks,'' \emph{IEEE
		Transactions on Mobile Computing}, vol.~13, no.~8, pp. 1774--1786, 2014.
	
	\bibitem{Patwari2014}
	N.~Patwari, L.~Brewer, Q.~Tate, O.~Kaltiokallio, and M.~Bocca, ``Breathfinding:
	A wireless network that monitors and locates breathing in a home,''
	\emph{IEEE Journal of Selected Topics in Signal Processing}, vol.~8, no.~1,
	pp. 30--42, Feb 2014.
	
	\bibitem{Rife1974}
	D.~Rife and R.~Boorstyn, ``Single tone parameter estimation from discrete-time
	observations,'' \emph{IEEE Transactions on Information Theory}, vol.~20,
	no.~5, pp. 591--598, 1974.
	
	\bibitem{Yigitler2017b}
	H.~Yi\u{g}itler, R.~J\"{a}ntti, O.~Kaltiokallio, and N.~Patwari, ``Detector
	based radio tomographic imaging,'' \emph{IEEE Transaction on Mobile
		Computing}, vol.~X, no.~X, pp. PP--PP, XX 2017.
	
	\bibitem{Kaltiokallio2017b}
	O.~Kaltiokallio, H.~Yi\u{g}itler, and R.~J\"{a}ntti, ``A three-state received
	signal strength model for device-free localization,'' \emph{IEEE Transaction
		on Vehicular Technology}, vol.~X, no.~X, pp. PP--PP, XX 2017.
	
	\bibitem{Hostettler2017}
	R.~Hostettler, O.~Kaltiokallio, H.~Yi\u{g}itler, S.~S\"{a}rkk\"{a}, and
	R.~J\"{a}ntti, ``{RSS}-based respiratory rate monitoring using periodic
	{Gaussian} processes and {Kalman} filtering,'' in \emph{25th European Signal
		Processing Conference (EUSIPCO)}, Kos, Greece, August 2017.
	
	\bibitem{Li2013}
	C.~Li, V.~M. Lubecke, O.~Boric-Lubecke, and J.~Lin, ``A review on recent
	advances in doppler radar sensors for noncontact healthcare monitoring,''
	\emph{IEEE Transactions on Microwave Theory and Techniques}, vol.~61, no.~5,
	pp. 2046--2060, May 2013.
	
	\bibitem{Salmi2012}
	J.~Salmi, O.~Luukkonen, and V.~Koivunen, ``Continuous wave radar based vital
	sign estimation: Modeling and experiments,'' in \emph{IEEE Radar Conference
		(RADAR), 2012}, 2012, pp. 0564--0569.
	
	\bibitem{Lazaro2010}
	A.~Lazaro, D.~Girbau, and R.~Villarino, ``Analysis of vital signs monitoring
	using an ir-uwb radar,'' \emph{Progress In Electromagnetics Research}, vol.
	100, pp. 265--284, 2010.
	
	\bibitem{Venkatesh2005}
	S.~Venkatesh, C.~R. Anderson, N.~V. Rivera, and R.~M. Buehrer, ``Implementation
	and analysis of respiration-rate estimation using impulse-based {UWB},'' in
	\emph{IEEE Military Communications Conference, 2005. MILCOM 2005.}\hskip 1em
	plus 0.5em minus 0.4em\relax IEEE, 2005, pp. 3314--3320.
	
	\bibitem{Adib2014}
	F.~Adib, Z.~Kabelac, D.~Katabi, and R.~C. Miller, ``{3D} tracking via body
	radio reflections.'' in \emph{NSDI}, vol.~14, 2014, pp. 317--329.
	
	\bibitem{Abdelnasser2015}
	H.~Abdelnasser, K.~A. Harras, and M.~Youssef, ``{UbiBreathe}: A ubiquitous
	non-invasive {WiFi}-based breathing estimator,'' in \emph{Proceedings of the
		16th ACM International Symposium on Mobile Ad Hoc Networking and
		Computing}.\hskip 1em plus 0.5em minus 0.4em\relax ACM, 2015, pp. 277--286.
	
	\bibitem{Liu2015}
	J.~Liu, Y.~Wang, Y.~Chen, J.~Yang, X.~Chen, and J.~Cheng, ``Tracking vital
	signs during sleep leveraging off-the-shelf {WiFi},'' in \emph{Proceedings of
		the 16th ACM International Symposium on Mobile Ad Hoc Networking and
		Computing}.\hskip 1em plus 0.5em minus 0.4em\relax ACM, 2015, pp. 267--276.
	
	\bibitem{Wang2016}
	H.~Wang, D.~Zhang, J.~Ma, Y.~Wang, Y.~Wang, D.~Wu, T.~Gu, and B.~Xie, ``Human
	respiration detection with commodity {WiFi} devices: do user location and
	body orientation matter?'' in \emph{Proceedings of the 2016 ACM International
		Joint Conference on Pervasive and Ubiquitous Computing}.\hskip 1em plus 0.5em
	minus 0.4em\relax ACM, 2016, pp. 25--36.
	
	\bibitem{Liu2016a}
	X.~Liu, J.~Cao, S.~Tang, J.~Wen, and P.~Guo, ``Contactless respiration
	monitoring via off-the-shelf {WiFi} devices,'' \emph{IEEE Transactions on
		Mobile Computing}, vol.~15, no.~10, pp. 2466--2479, 2016.
	
	\bibitem{Luong2016}
	A.~Luong, A.~S. Abrar, T.~Schmid, and N.~Patwari, ``{RSS} step size: 1 {dB} is
	not enough!'' in \emph{Proceedings of the 3rd Workshop on Hot Topics in
		Wireless}.\hskip 1em plus 0.5em minus 0.4em\relax ACM, 2016, pp. 17--21.
	
	\bibitem{Yigitler2017}
	H.~Yi\u{g}itler, R.~J\"{a}ntti, and N.~Patwari, ``On log-normality of {RSSI} in
	narrowband receivers under static conditions,'' \emph{IEEE Signal Processing
		Letters}, vol.~24, no.~4, pp. 367--371, April 2017.
	
	\bibitem{Rappaport2002}
	T.~S. Rappaport, \emph{Wireless Communications: Principles and Practice},
	2nd~ed.\hskip 1em plus 0.5em minus 0.4em\relax Prentice Hall, 2002.
	
	\bibitem{Carlson2002}
	A.~B. Carlson, P.~B. Crilly, and J.~Ruttledge, \emph{Communication systems: An
		introduction to signals and noise in electrical communication}, 4th~ed.\hskip
	1em plus 0.5em minus 0.4em\relax McGraw-Hill New York, 2002.
	
	\bibitem{Abramowitz1970}
	M.~Abramowitz and I.~A. Stegun, \emph{Handbook of mathematical functions: with
		formulas, graphs, and mathematical tables}.\hskip 1em plus 0.5em minus
	0.4em\relax Dover Publications, 1970.
	
	\bibitem{Kaneko2012}
	H.~Kaneko and J.~Horie, ``Breathing movements of the chest and abdominal wall
	in healthy subjects,'' \emph{Respiratory Care}, vol.~57, no.~9, pp.
	1442--1451, 2012.
	
	\bibitem{Maximon2003}
	L.~C. Maximon, ``The dilogarithm function for complex argument,''
	\emph{Proceedings of the Royal Society of London. Series A: Mathematical,
		Physical and Engineering Sciences}, vol. 459, no. 2039, pp. 2807--2819, 2003.
	
	\bibitem{Kay1988}
	S.~M. Kay, \emph{Modern Spectral Estimation}.\hskip 1em plus 0.5em minus
	0.4em\relax Prentice-Hall, 1988.
	
	\bibitem{Qi2002}
	Y.~Qi, T.~P. Minka, and R.~W. Picard, ``{Bayesian} spectrum estimation of
	unevenly sampled nonstationary data,'' in \emph{IEEE International Conference
		on Acoustics, Speech, and Signal Processing (ICASSP)}, vol.~2, May 2002, pp.
	1473--1476.
	
	\bibitem{Sarkka2013}
	S.~S\"{a}rkk\"{a}, A.~Solin, and J.~Hartikainen, ``Spatiotemporal learning via
	infinite-dimensional {Bayesian} filtering and smoothing: A look at {Gaussian}
	process regression through {Kalman} filtering,'' \emph{IEEE Signal Processing
		Magazine}, vol.~30, no.~4, pp. 51--61, July 2013.
	
	\bibitem{Rasmussen2006}
	C.~E. Rasmussen and C.~K.~I. Williams, \emph{Gaussian Processes for Machine
		Learning}.\hskip 1em plus 0.5em minus 0.4em\relax MIT Press, 2006.
	
	\bibitem{Solin2014}
	A.~Solin and S.~S\"arkk\"a, ``Explicit link between periodic covariance
	functions and state space models,'' in \emph{Proceedings of the 17th
		International Conference on Artificial Intelligence and Statistics
		({AISTATS})}, vol.~33, 2014, pp. 904--912.
	
	\bibitem{Oksendal2010}
	B.~{\O}ksendal, \emph{Stochastic Differential Equations: An Introduction with
		Applications}, 6th~ed.\hskip 1em plus 0.5em minus 0.4em\relax Springer, 2010.
	
	\bibitem{Yigitler2014}
	H.~Yi\u{g}itler, R.~J\"{a}ntti, and R.~Virrankoski, ``{pRoot}: An adaptable
	wireless sensor-actuator hardware platform,'' in \emph{12th IEEE
		International Conference on Embedded and Ubiquitous Computing (EUC)}.\hskip
	1em plus 0.5em minus 0.4em\relax IEEE, 2014, pp. 281--286.
	
\end{thebibliography}

\end{document}